\documentstyle[psfig]{l-aa}
\newcommand{\be}{\begin{equation}}
\newcommand{\ee}{\end{equation}}
\newcommand{\bea}{\begin{eqnarray}}
\newcommand{\eea}{\end{eqnarray}}

\newcommand{\msun}{M_\odot}
\newcommand{\rsun}{R_\odot}
\newcommand{\lsun}{L_\odot}
\newcommand{\ca}[2]{{\cal {#1}}_{#2}}
\newcommand{\zs}{\zeta_{\scriptscriptstyle s}}
\newcommand{\zeq}{\zeta_{\scriptscriptstyle eq}}
\newcommand{\zl}{\zeta_{\scriptscriptstyle L}}
\newcommand{\zlr}{\zeta_{\scriptscriptstyle L,ring}}
\newcommand{\rl}{r_{\scriptscriptstyle L}}
\newcommand{\rloa}{(\rl/a)}
\newcommand{\mi}{m_{\scriptscriptstyle 1}}
\newcommand{\mio}{m_{\scriptscriptstyle 1,0}}
\newcommand{\mii}{m_{\scriptscriptstyle 2}}
\newcommand{\miio}{m_{\scriptscriptstyle 2,0}}
\newcommand{\rii}{r_{\scriptscriptstyle 2}}
\newcommand{\mc}{m_{\scriptscriptstyle c}}
\newcommand{\me}{m_{\scriptscriptstyle e}}
\newcommand{\mg}{m_{\scriptscriptstyle g}}
\newcommand{\mx}{m_{\scriptscriptstyle X}}
\newcommand{\mxed}{\dot m_{\scriptscriptstyle X,Edd}}
\newcommand{\mt}{M_{\scriptscriptstyle T}}
\newcommand{\mto}{M_{\scriptscriptstyle T,0}}
\newcommand{\cno}{\varepsilon_{\scriptscriptstyle {CNO}}}
\newcommand{\dln}[1]{\partial \ln {#1}}
\newcommand{\zshw}{\zeta_{\scriptscriptstyle HW}}
\newcommand{\zsspho}{\zeta_{\scriptscriptstyle SPH1}}
\newcommand{\zssph}{\zeta_{\scriptscriptstyle SPH}}
\newcommand{\ew}{\epsilon_{\scriptscriptstyle w}}
\newcommand{\er}{\epsilon_{\scriptscriptstyle r}}

\begin{document}
%\begin{document}
%
   \thesaurus{06       % A&A Section 6: Form. struct. and evolut. of stars
            (08.07.02;  % Stars: giant
             08.02.03;  % (Stars:) binaries: general
             08.02.05;  % (Stars:) binaries: symbiotic
	     08.14.01;  % Stars: neutron
	     08.14.02;  % novae, cv
	     08.16.06   % pulsars: general
%           ;  08.16.4;  % Stars: asymptotic, post-asymptotic giant branch
%             10.15.1 % (Galaxy:) open clusters and associations: general
	     )}
\title{Stability Criteria for Mass Transfer in Binary Stellar Evolution}

\author{G.\,E.\, Soberman\inst{1},
	E.\,S.\, Phinney\inst{1},
	\and
	E.\,P.\,J.\, van den Heuvel\inst{2}
        }

\offprints{G. E. Soberman; Internet Address: {\tt soberman@tapir.caltech.EDU} }

\institute{Theoretical Astrophysics,
           130-33 California Institute of Technology,
           Pasadena, CA 91125
           USA
          \and 
	   Astronomical Institute and Center for High Energy Astrophysics (CHEAF),
           University of Amsterdam,
	   Kruislaan 403,
	   1098 SJ Amsterdam
           The Netherlands
          }

\date{Received December, 1996; accepted February 1997.}

\maketitle

\markboth{G.\,E.\,Soberman, E.\,S.\, Phinney \&\ E.\,P.\,J.\, van den
Heuvel }{Stability Criteria for Mass Transfer in Binary Stellar Evolution}

\begin{abstract}
The evolution of a binary star system by various analytic
approximations of mass transfer is discussed, with particular
attention payed to the stability of these processes against runaway on
the thermal and dynamical timescales of the mass donating star.  
Mass transfer in red giant - neutron star binary systems is used as
a specific example of such mass transfer, and is investigated.
Hjellming and Webbink's (\cite{hw87}) results on the dynamic
timescale response of a convective star with a core to mass loss are
applied, with new results.

It is found that mass transfer is usually stable, so long as the the
wind's specific angular momentum does not exceed the angular momentum
per reduced mass of the system. This holds for both dynamical and
thermal timescales.  Those systems which are not stable will usually
transfer mass on the thermal timescale.  Included are graphs
illustrating the variation of $\frac{\dln\rl}{\dln m} \equiv \zl$ with
mass ratio in the binary, for various parameters in the
non-conservative mass transfer, as well as evolutionary paths of
interacting red giant neutron star binaries.
%
%The second appendix of this paper, describing extenstions to models
%presented in the body, is available in electronic form.


\end{abstract}

\section{Introduction}

The dominant feature in the evolution of stars in tight binaries, and
the one which distinguishes it from the evolution of single stars is
the presence of various forms of mass transfer between the two stars.
Mass transfer occurs in many different types of systems, to widely
varying effects (cf.: Shore \cite{shore}): contamination of the
envelope of a less evolved star with chemically processed elements, as
in Barium stars; winds from one star, which may be visible as a screen
in front of the other, as in EM~Car, AO~Cas, and most notably the
binary PSR 1259-63 (Johnston, et al., \cite{johnston92};
Kochanek, \cite{kochanek}; Melatos, et al., \cite{melatos};
Johnston, et al., \cite{johnston95}); catastrophic mass
transfer, by common envelope phase, as in W~UMa systems; or a slow,
steady mass transfer by Roche lobe overflow.

Mass transfer is particularly interesting if one considers the
evolution of a system with at least one degenerate star.  In these
cases, mass transfer produces spectacular effects, resulting in part from
the intense magnetic and gravitational fields of the compact
objects -- pulsed X-ray emission, nuclear burning, novae outbursts,
and so on.  Also, since the mass transfer rates can be high, and
orbital period measurements accurate, one may see the dynamical
effects of mass transfer on the binary orbit, as in Cygnus X-3 (van
Kerkwijk, {\it et.al.}, \cite{vankerkwijk}; van den Heuvel \cite{vdh94}).

In the case of cataclysmic variables (CVs) and low mass X-ray binaries
(LMXBs), one has a highly evolved, compact star (CVs and LMXBs contain
white dwarfs and neutron stars, respectively) and a less evolved main
sequence or red giant star.  Mass transfer usually proceeds by
accretion onto the compact object, and is secularly stable.   
The
transfer may be accompanied by a stellar wind from the mass-losing
star, or ejection of matter from the accretor, as in novae and
galactic jet sources.

There are various unanswered questions in the evolution of LMXBs into
low mass binary pulsars (LMBPs, in which a millisecond radio pulsar is
in a binary with a low mass white dwarf companion) and of CVs.  Among
these are the problem of the disparate birthrates of the LMXBs and the
LMBPs (Kulkarni and Narayan \cite{nk}), estimation of the strength of
X-ray heating induced winds from the donor star (London, {\it et.al.}
\cite{lond81}; London and Flannery \cite{lond82}; Tavani and
London \cite{tav93}; Banit and Shaham \cite{bs}), effects of X-ray
heating on the red giant structure (Podsiadlowski \cite{podsinat};
Harpaz and Rappaport \cite{hr}; Frank,{\it et.al.} \cite{fkl};
Hameury,{\it et.al.} \cite{hklr}),
{\it etc.}

The problem of mass transfer in binaries by Roche lobe overflow has
received a good deal of attention in the literature over the past few
decades, typically in investigations of one aspect or another of
orbital evolution or stability.  Ruderman, {\it et. al.}
(\cite{rudshatav}) examined mass transfer by isotropic winds and
accretion, in investigating the evolution of short period binary X-ray
sources with extreme mass ratios, such as PSR~1957+20 and 4U1820-30.
King and Kolb (\cite{KK}) developed models with accretion and
(typically) isotropic re-emission of transferred matter, in the
context of the period gap in cataclysmic variables.

The aim of this paper is to present a unified treatmant of binary
orbital evolution and stability, with relevant equations in a general
form.  The limits of pure modes of mass transfer and extreme mass
ratios are presented and examined, to gain a qualitative understanding
of slow, non-disruptive mass transfer in its various forms.

The paper is organised as follows.  Kinematic equations for orbital
evolution, assuming mass transfer through some well-specified mode(s),
are derived in Sect.~(\ref{secmx}).  Conditions for and regions of
stability are derived in Sect.~(\ref{zetas}).  The theory is applied
to LMXBs in Sect.~(\ref{rgns}), and conclusions drawn in the final
section.  Two appendices contain further details on the unified models
and on comparisons among the various limits of the
models
% \footnote{Appendix \ref{hfpf} (``Extended Models'') is available via
% anonymous ftp from {\tt cdsarc.u-strasbg.fr} (130.79.128.5) or at {\tt
% http://cdsweb.u-strasbg.fr/Abstract.html}}
.

\section{Mass Transfer and the Evolution of Orbital Parameters}
\label{secmx}

In this section, expressions for the variation of orbital parameters
with loss of mass from one of the stars are derived.  In what follows,
the two stars will be referred to as $\mi$ and $\mii$, with the latter
the mass losing star.

A binary composed of two stars with radii of gyration much less than
the semimajor axis $a$ will have an angular momentum: 
\be
L = \mu \frac{2\pi}{P}a^2(1-e^2)^{1/2} \; , \label{L}
\ee

\noindent where the period $P$ is related to $a$ and the total mass
$\mt = \mi + \mii$ through Kepler's law:

\be 
G\mt = a^3\left(\frac{2\pi}{P} \right)^2 \; . \label{kep} 
\ee
\noindent Here, the reduced mass $\mu = \mi\mii/\mt$ and the mass
ratio $q= \mii/\mi$.  Tidal forces circularise the orbits of
semidetatched binaries on timescales of $10^4 \mbox{y} \ll
\tau_{nuc}$,\footnote{The tidal timescale is also much shorter than
the timescale of angular momentum loss by gravitational wave
radiation, so binaries in which mass transfer is driven by angular
momentum loss are also tidally locked.}  the nuclear timescale on
which the binaries evolve (Verbunt and Phinney \cite{vp95}).
Furthermore, one expects stable mass transfer by RLOF to circularise
orbits.  Consequently, an eccentricity $e=0$ will be assumed for the
remainder of this work.

Equations (\ref{L}) and (\ref{kep}) combine to give expressions for the
semimajor axis and orbital period, in terms of the masses and angular
momentum:

\bea
a & = & \left(\frac{L}{\mu} \right)^2 (G\mt)^{-1} \; , \label{a} \\
\frac{P}{2\pi} & = & \left(\frac{L}{\mu} \right)^3 (G\mt)^{-2} \; , \label{P}
\eea

\noindent with logarithmic derivatives:

\bea
\dln a & = & 2\,\dln (L/\mu) - \dln \mt \; , \label{da} \\
\dln P & = & 3\,\dln (L/\mu) - 2\,\dln \mt \; . \label{dP}
\eea

There are various modes (or, borrowing from the terminology of nuclear
physics, channels) of mass transfer associated with the different
paths taken by, and destinations of mass lost from $\mii$.

The details of mass transfer must be considered, in order to calculate
the orbital evolution appropriate to each mode.  Several of these
modes are described here, roughly following Section 2.3 of the review
article by van den Heuvel (\cite{vdh94}).  

The mode which is most
often considered is accretion, in which matter from $\mii$ is
deposited onto $\mi$.
The accreted component (slow mode) conserves total mass $\mt$ and
orbital angular momentum $L$.  In the case considered here, of a Roche
lobe filling donor star, matter is lost from the vicinity of the donor
star, through the inner Lagrangian point, to the vicinity of the
accretor, about which it arrives with a high specific angular
momentum.  If the mass transfer rate is sufficiently high, an
accretion disk will form ({\it c.f.:} Frank, King, and
Raine,~\cite{FKR}).  The disk forms due to viscosity of the proferred
fluid and transports angular
momentum away from and mass towards the accretor.  As angular momentum
is transported outwards, the disk is expands to larger and
larger circumstellar radii, until significant tides develop between
the disk and the mass donor.  These tides transfer angular momentum
from the disk back to the binary and inhibit further disk growth (Lin and
Papaloizou \cite{linpap}).  The mass transfer process is conservative
if all mass lost from the donor ($\mii$) is accreted in this way.

The second mode considered is Jeans's mode, which van den Heuvel
calls, after Huang (\cite{huang}), the fast mode; the third is
isotropic re-emission.  Jeans's mode is a spherically symmetric
outflow from the donor star in the form of a fast wind.  The best
known example of Jeans's mode: the orbital evolution associated with
Type II supernovae in binaries (Blaauw, \cite{blaauw}) differs
markedly from what is examined here.  In that case, mass loss is
instantaneous.  A dynamically adiabatic outflow ($m/\dot{m} \gg
P_{\scriptscriptstyle orb}$) is considered here.

An interesting variant of Jeans's mode is isotropic re-emission.
This is a flow in which matter is transported from the donor star to
the vicinity of the accretor, where it is then ejected as a fast,
isotropic wind.  The distinction between Jeans's mode and isotropic
re-emission is important in considerations of angular momentum loss
from the binary.  Mass lost {\it via} the the Jeans mode carries the
specific angular momentum of the mass loser; isotropically
re-emitted matter has the specific angular momentum of the
accretor.

Wind hydrodynamics are ignored.

The fourth case considered is the intermediate mode, or mass loss to a
ring.  No mechanism for mass loss is hypothesised for this mode. The
idea is simply that the ejecta has the high specific angular momentum
characteristic of a circumbinary ring.

The differences among each of these modes is in the variation of the
quantities $q$, $\mt$, and most importantly $L/\mu$, with mass loss.
For example, given a $q=0.25$ binary, mass lost into a circumbinary
ring ($a_r>a$) removes angular momentum at least 25 times faster
than mass lost by isotropic re-emission.  This difference will have
a great effect on the evolution of the orbit, as well as on the
stability of the mass transfer.

In the following sections, formulae for the orbital evolution of a
binary under different modes of mass loss are presented and integrated
to give expressions for the change in orbital parameters with mass
transfer.  The different modes are summarised in
Table~(\ref{modenames}).

Due to the formality of the following section and the cumbersome
nature of the formulae derived, one should try to develop some kind of
intuitition, and understanding of the results.  The two sets of
results ({\it i.e.:} one for a combination of winds, the other for
ring formation) are examined in the limits of extreme mass ratios,
compared to one another, and their differences reconciled in the first
Appendix.

\begin{table}
\begin{tabular}{llcc}
Name of Mode& Type & $-\frac{\partial m_{mode}}{\partial\mii}$
&$\mu h_{loss}/L$ \\
\vspace{-2mm} \\ \hline
\vspace{-2mm} \\
Jeans/Fast&Isotropic wind&$\alpha$& $\frac{A}{(1+q)^2}$ \\
\multicolumn{2}{c}{--- Isotropic Re-emission ---}&$\beta$&
$\frac{q}{(1+q)^2}$ \\
Intermediate&Ring&$\delta$&$\gamma$ \\
Slow &Accretion& $\epsilon$&$0$
\end{tabular}
\label{modenames}
%\caption []{Names and brief description of the modes of mass loss
%explored in this paper.  The factor $A$ in the fast mode is typically
%unity.  See Sect.~(\ref{unified}) for details.}
\caption []{Names, brief description of, and dimensionless specific
angular momentum of the modes of mass loss explored in this paper.
Parameters $\alpha$, $\beta$, $\gamma$, $\delta$, $\epsilon$, and $A$
are defined in Eqs. (\ref{adef}), (\ref{bdef}), (\ref{hr}),
(\ref{dLdmr1}), and (\ref{dLdmw2}), respectively.  The factor $A$ in
the fast mode is typically unity.}
\end{table}

\subsection{Unified Wind Model} \label{unified}

Consider first a model of mass loss which includes isotropic wind,
isotropic re-emission, and accretion, in varying strengths.
Conservative mass transfer, as well as pure forms of the two isotropic
winds described above, can then be recovered as limiting cases by
adjusting two parameters of this model:

\bea
\alpha & = & - \frac{\partial m_{wind}}{\partial \mii} \; , \label{adef} \\
\beta & = & - \frac{\partial m_{iso-r}}{\partial \mii} \; . \label{bdef}
\eea

\noindent Here, $\partial m_{wind}$ is a small mass of wind from
star 2 (isotropic wind), and $\partial m_{iso-r}$ a small mass of wind from
star 1 (isotropic re-emission).  Both amounts are expresses in
terms of the total amount of mass $\partial\mii$ lost from star 2.
A sign convention is chosen such
that $\partial m_{star}$ is positive if the star's mass increases, and
$\partial m_{flow}$ is positive if removing matter (from $\mii$).
Given the above, one can write formulae for the variation of all the
masses in the problem:

\bea
\ew & \equiv & 1- \alpha - \beta \label{edefw} \\
\dln \mi & = & -q\ew\, \dln\mii \label{dmidmw} \\
\dln \mt & = & \frac{q}{1+q}(\alpha + \beta )\,  \dln \mii \label{dmtdmw} \\
\dln q & = & (1 + \ew q)\, \dln \mii \label{dqdmw} \\
\vspace{2.0mm} \nonumber \\
\dln \mu & = & \dln \mii  - \frac{q}{1+q}\dln q \label{dmudmw} \; .
\eea

Neglecting accretion onto the stars from the ISM and mass currents
originating directly from the accretor, all transferred mass comes
from the mass donating star ($\mii$) and $\alpha$, $\beta$, and the
accreted fraction $\ew$ all lie between $0$ and $1$, with the
condition of mass conservation $\alpha + \beta + \ew = 1$, imposed by
Eq.~(\ref{edefw}).  For the remainder of this section, the subscript
on $\ew$ will be eschewed.

If mass is lost isotropically from a nonrotating star, it carries no
angular momentum in that star's rest frame.  In the center of mass
frame, orbital angular momentum $L$ will be removed at a rate $\dot L
= h\,\dot m$, where $\dot m$ is the mass loss rate and $h$ the
specific angular momentum of the orbit.

If the windy star is rotating, then it loses spin angular momentum
$S$ at a rate $\dot S = R_W^2 \Omega_* \dot m$, where $\Omega_*$ is
that star's rotation rate and $R_W^2$ the average of the square of the
perpendicular radius at which the wind decouples from the star.
Strongly magnetised stars with ionised winds will have a wind
decoupling radius similar to their Alfv\'{e}n radius, and spin angular
momentum may be removed at a substantially enhanced rate.

The above consideration becomes important in the case of a tidally
locked star, such as a Roche lobe filling red giant.  In this case, if
magnetic braking removes spin angular momentum from the star at some
enhanced rate, that star will begin to spin asynchronously to the
orbit and the companion will establish tides to enforce corotation.
These tides are of the same form as those between the oceans and the
Moon, which are forcing the length of the day to tend towards that of
the month.

Strong spin-orbit coupling changes the evolution of orbital angular
momentum in two ways.  First, for a given orbital frequency, there is
an extra store of angular momentum, due to the inertia of the
star. 
% Thus, for a given torque, $L$ evolves more slowly, as ...
% is a
% slowing of the loss of angular momentum, due to having to torque the
% star about its axis, as well as about its orbit.  
Thus, for a given torque, $L$ evolves more slowly, by a factor $L/(L+S)$.
The second is from
the enhanced rate of loss of total angular momentum, due to the loss
of spin angular momentum from the star.  Extreme values of $R_W/R_*$
allow the timescale for loss of spin angular momentum to be much less
than that for orbital angular momentum.  Thus, the second effect will
either compensate for or dominate over the first, and there will be an
overall increase in the torque due to this wind.

This enhancement will be treated formally, by taking the angular momentum
loss due to the fast mode to occur at a rate $A$ times greater than
what would be obtained neglecting the effects of the finite sized
companion.  

Keeping the above discussion in mind, the angular momentum lost from
the system, due to winds is as follows:
\be
\partial L = (A\alpha h_2 +\beta h_1)\partial\mii \label{dLdmw2}\, , 
\ee
\noindent which can be simplified by substituting in expressions for
the $h_i$:
\bea
h_i & = & L_i/m_i \nonumber \\ 
& = & (L/\mu)\frac{\mt-m_i}{\mt}\frac{\mu}{m_i} \label{hi}\\ 
h_1 & = & \frac{q^2}{1+q}\frac{L}{\mii} \label{h1} \\ 
h_2 & = & \frac{1}{1+q}\frac{L}{\mii} \label{h2} \\
\partial L & = & \frac{A\alpha + \beta q^2}{1+q}
\frac{L}{\mii} \partial\mii \label{dLdmw1}\\
\dln L & = &\frac{A\alpha + \beta q^2}{1+q} \dln\mii \, .\label{dLdmw}
\eea

So far, the equations have been completely general; $\alpha$, $\beta$,
and $A$ may be any functions of the orbital elements and stellar
properties.  Restricting these functions to certain forms leads to simple
integrable models of orbital evolution.  If constant fractions of the
transferred mass pass through each channel ($\alpha$, $\beta$
constant), then the masses are expressable as simple functions of the
mass ratio $q$:
\bea
\frac{\mii}{\miio} & = & 
\left( \frac{q}{q_0} \right)
\left( \frac{1+\epsilon q_0}{1+\epsilon q}  \right)
\label{miiw} \\
\frac{\mi}{\mio} & = & 
\left( \frac{1+\epsilon q_0}{1+\epsilon q}  \right)
\label{miw} \\
\frac{\mt}{\mto} & = & 
\left( \frac{1+q}{1+q_0} \right)
\left( \frac{1+\epsilon q_0}{1+\epsilon q}  \right)
\label{mtw} \\
\frac{\mu}{\mu_0} & = & 
\left( \frac{q}{q_0} \right)
\left( \frac{1+q_0}{1+q}  \right)
\left( \frac{1+\epsilon q_0}{1+\epsilon q}  \right)
\label{muw}
\eea

\noindent Furthermore, if the enhancement factor $A$ is also
constant\footnote{This may not be the best approximation, if the orbit
widens or narrows significantly and the torque is dominated by magnetic
braking.}, then the angular momentum is an integrable function of $q$,
as well:

\be
\frac{L}{L_0} = \left(\frac{q}{q_0}\right)^{\ca{A}{w}}
\left(\frac{1+q_0}{1+q}\right)^{\ca{B}{w}}
\left(\frac{1+\epsilon q}{1+\epsilon q_0}\right)^{\ca{C}{w}}
\,, \label{Lw}
\ee
\noindent where the exponents\footnote{There is no problem when the
denominator $1-\epsilon \rightarrow 0$, as the numerator vanishes at
the same rate.  For $\ca{C}{w}$, the power laws become exponentials in
the absence of accretion ($\epsilon=0$). } are given by:

\bea
\ca{A}{w} & = & A\alpha \,, \label{qexpw} \\ 
\ca{B}{w} & = & \frac{A\alpha + \beta}{1-\epsilon} \,, \label{qpoexpw} \\ 
\ca{C}{w} & = & \frac{A\alpha\epsilon}{1-\epsilon}+
\frac{\beta}{\epsilon(1-\epsilon)} \,. \label{eqpoexpw}
\eea

Finally, substitution of Eqs.~(\ref{mtw}), (\ref{muw}), and (\ref{Lw}) into
Eqs.~(\ref{a}) and (\ref{P}) gives expressions for the evolution of
the semimajor axis and orbital period in terms of the changing mass
ratio $q$:

% winds, only:
\bea
\frac{a}{a_0} & = & 
\left(\frac{q}{q_0}\right)^{2\ca{A}{w}-2}
\left(\frac{1+q}{1+q_0}\right)^{1-2\ca{B}{w} } \nonumber \\
& & \left(\frac{1+\epsilon q}{1+\epsilon q_0}\right)^{3+2\ca{C}{w}} \, ,
\label{aw} \\
\frac{P}{P_0} & = & 
\left(\frac{q}{q_0}\right)^{3\ca{A}{w}-3}
\left(\frac{1+q}{1+q_0}\right)^{1-3\ca{B}{w} } \nonumber \\
& & \left(\frac{1+\epsilon q}{1+\epsilon q_0}\right)^{5+3\ca{C}{w}} \, .
\label{Pw}
\eea

The derivatives of these functions may be evaluated either by
logarithmic differentiation of the above expressions (Eqs.~(\ref{aw})
and (\ref{Pw})), or by substitution of Eqs.~(\ref{dmtdmw}),
(\ref{dqdmw}), (\ref{dmudmw}), and (\ref{dLdmw}) into Eqs.~(\ref{da}) and
(\ref{dP}).  Either way, the results are:

\bea
\frac{\dln a}{\dln q} & = & 2(\ca{A}{w}-1) + 
(1-2\ca{B}{w})\frac{q}{1+q} \nonumber \\ 
& & + (3+2\ca{C}{w}) \frac{q}{1+\epsilon q}
\, , \label{daw} \\
\frac{\dln P}{\dln q} & = & 3(\ca{A}{w}-1) + 
(1-3\ca{B}{w})\frac{q}{1+q} \nonumber \\ 
& & + (5+3\ca{C}{w})\frac{q}{1+\epsilon q} 
\, .\label{dPw} 
\eea

The reader should be convinced that the above equations are correct.
First, and by fiat, they combine to give Kepler's law.  Second, they
reduce to the familiar conservative results:

% conservative case:
\bea
\frac{a}{a_0} & = & \left(\frac{q_0}{q}\right)^2
	\left(\frac{1+q}{1+q_0}\right)^4 \; , \label{ac} \\
\frac{P}{P_0} & = & \left(\frac{q_0}{q}\right)^3
	\left(\frac{1+q}{1+q_0}\right)^6  \; , \label{Pc}
\eea

\noindent in the $\epsilon=1$ limit.  Finally, and again by
construction, the formulae are composable.  Ratios such as $P/P_0$
are all of the functional form $f(q)/f(q_0)$, so if one forms 
e.g.:  $(P_2/P_0) = (P_2/P_1)(P_1/P_0)$, the result is immediately
independent of the arbitrarily chosen intermediate point $P_1 =
P(q_1)$.

Note that the results for isotropic re-emission obtained by
Bhattacharya and van den Heuvel (\cite{vdh91}, Eq.~(A.6)), and again
by van den Heuvel (\cite{vdh94}, Eq.~(40)) are not composable, in the
sense described above.  The correct expression for $(a/a_0)$ in the
case of isotropic re-emission may be obtained by setting $\alpha =
0$:

\bea
\frac{a}{a_0} & = & 
\left(\frac{q_0}{q}\right)^2
\left(\frac{1+q_0}{1+q}\right) 
\left(\frac{1+(1-\beta)q}{1+(1-\beta)q_0}\right)
^{5 + \frac{2\beta}{1-\beta}} \label{air}.
\eea

\noindent Also note that Eq.~(A.7) of Bhattacharya and van den Heuvel
(\cite{vdh91}) has its exponential in the denominator; it should be
in the numerator. In van den Heuvel (\cite{vdh94}), Eq.~(37) should
read $\dln J = \frac{\beta q^2}{1+q} \frac{\dln q}{1+(1-\beta)q}$.  In
Eq.~(38), an equals sign should replace the plus.  (These corrections
were also found by Tauris (\cite{tauris})).

Thus, if there is mass loss from one star, with constant fractions of
the mass going into isotropic winds from the donor and its companion,
one can express the variation in the binary parameters, $\mt$, $a$,
and $P$, in terms of their initial values, the initial and final
values of the ratio of masses, and these mass fractions.

\subsection{Formation of a Coplanar Ring} \label{ring}

Now consider a model in which mass is transferred by accretion and
ring formation, as described above.  For concreteness, follow a
standard prescription ({\it c.f.:} van den Heuvel (\cite{vdh94})) and
take the ring's radius, $a_{r}$ to be a constant multiple $\gamma^2$
of the binary semimajor axis.  This effectively sets the angular
momentum of the ring material, since for a light ring, 

\be
h_{r} = L_r/m_r = \sqrt{G\mt a_r} = \gamma L/\mu\,. \label{hr}
\ee

Formulae describing orbital evolution can be obtained acording to the
prescription of the previous section.  If a fraction $\delta$ of the
mass lost from $\mii$ is used in the formation of a ring, then Eqs.
(\ref{edefw}) and (\ref{dLdmw}) should be replaced as follows:

\bea
\er & \equiv & 1- \delta \label{edefr} \\
\partial L & = & \delta h_r \, \partial\mii \nonumber\\
& = & \gamma\delta L/\mu  \, \partial\mii \nonumber\\
& = & \gamma\delta(1+q) L/\mii \, \partial\mii \label{dLdmr1}\\
\dln L & = & \gamma\delta(1+q) \,\dln\mii\,. \label{dLdmr}
\eea

Following the procedure used in Sect.~(\ref{unified}), and with
$\delta=1$ ($\er=0$):

\bea
\frac{\mt}{\mto} & = & \left(\frac{1+q}{1+q_0}\right) 
	\; , \label{mtr} \\
\vspace{2mm}\nonumber\\
\frac{L}{L_0} & = & \left(\frac{q}{q_0}\right)^{\gamma}
\exp(\gamma(q-q_0)) \; , \label{Lr} \\
\vspace{2mm}\nonumber\\
\frac{a}{a_0} & = & \left(\frac{q}{q_0}\right)^{2(\gamma-1)}
	\left(\frac{1+q}{1+q_0}\right) \exp{(2\gamma(q-q_0))}
	\; , \label{ar} \\
\vspace{2mm}\nonumber\\
\frac{P}{P_0} & = & \left(\frac{q}{q_0}\right)^{3(\gamma-1)}
	\left(\frac{1+q}{1+q_0}\right) \exp(3\gamma(q-q_0))
	\; . \label{Pr} 
\eea

\bea
\frac{\dln a}{\dln q} & = & 2\gamma(1+q)-2+\frac{q}{1+q}\, , \label{dar} \\
\frac{\dln P}{\dln q} & = & 3\gamma(1+q)-3+\frac{q}{1+q}\, . \label{dPr}
\eea

\noindent Unless the ring is sufficiently wide ($a_r/a \ga 2$), it
will orbit in a rather uneven potential, with time-dependant tidal
forces which are comparable to the central force.  In such a
potential, it would likely fragment, and could fall back upon the
binary.  Consequently, stability probably requires the ring to be at a
radius of at least a few times $a$.  A `bare-minimum' for the ring
radius is the radius of gyration of the outermost Lagrange point of
the binary is between $a$ and $1.25a$, depending on the mass ratio of
the binary (Pennington, \cite{penn85}).  The ring should not sample
the potential at this radius.  For most of what follows, we will work
with a slightly wider ring: $a_r/a = \gamma^2=2.25$ and $\delta=1$.

\section{Linear Stability Analysis of the Mass Transfer} \label{zetas}

Mass transfer will proceed on a timescale which depends critically on
the changes in the radius of the donor star and that of its Roche lobe
in response to the mass loss.  The mass transfer might proceed on the
timescale at which the mass transfer was initially driven (e.g.:
nuclear, or orbital evolutionary), or at one of two much higher
rates: dynamical and thermal.

If a star is perturbed by removal of mass, it will fall out of
hydrostatic and thermal equilibria, which will be re-established on
sound crossing (dynamical) and heat diffusion (Kelvin-Helmholtz, or
thermal) timescales, respectively.  As part of the process of
returning to equilibrium, the star will either grow or shrink, first
on the dynamical, and then on the (slower) thermal timescale.  At the
same time, the Roche lobe also grows or shrinks around the star in
response to the mass loss.  If after a transfer of a small amount of
mass, the star's Roche lobe continues to enclose the star, then the
mass transfer is stable, and proceeds on the original driving
timescale.  Otherwise, it is unstable and proceeds on the fastest
unstable timescale.

In stability analysis, one starts with the equilibrium situation and
examines the small perturbations about it.  In this case, the question
is whether or not a star is contained by its Roche lobe.  Thus, one
studies the behaviour of the quantity
\be
\Delta\zeta = \frac{m}{R} \, \frac{\delta \Delta R}{\delta m}\; , 
\ee
which is the (dimensionless) variation in the difference in radius
between the star and its Roche lobe, in response to change in that
star's mass.  Here $\Delta R$ is the difference between the stellar
radius $R_*$ and the volume-equivalent Roche radius $\rl$.  The star
responds to this loss of 
%perturbation in
mass on two widely separated different timescales, so this analysis
must be performed on both of these timescales.

The linear stability analysis then amounts to a comparison of the
exponents in power-law fits of radius to mass, $R \sim m^{\zeta}$:

\bea
\zs & = & \left. \frac{\dln R_*}{\dln m} 
                 \right|_s  \; , \label{zs} \\
\zeq & = & \left. \frac{\dln R_*}{\dln m} 
                 \right|_{eq}  \; , \label{zeq} \\
\zl & = & \left. \frac{\dln \rl}{\dln m}
                 \right|_{bin. evol.} \; , \label{zl} 
\eea

\noindent where $R_*$ and $m$ refer to the mass-losing, secondary
star.  Thus, $R_*=\rii$ and $m=\mii$.  Stability requires that after
mass loss ($\delta\mii<0$) the star is still contained by its Roche
lobe.  Assuming $\Delta R_2 =0$ prior to mass loss, the stability
condition then becomes $\delta\Delta R_2 < 0$, or $\zl < (\zs, \zeq)$.
If this is not satisfied, then mass transfer runs to the fastest,
unstable timescale.

Each of the exponents is evaluated in a manner consistent with the
physical process involved.  For $\zs$, chemical abundance and entropy
profiles are assumed constant and mass is removed from the outside of
the star.  For $\zeq$, mass is still removed from the outside of the
star, but the star is assumed to be in the thermal equilibrium state
for the given chemical profile.  In calculating $\zl$, derivatives are
to be taken along the assumed evolutionary path of the binary system.

In the following subsections these exponents are described a bit
further and computed in the case of mass loss from a binary containing
a neutron star and a Roche lobe-filling red giant.  Such systems are
thought to be the progenitors of the wide orbit, millisecond pulsar,
helium white dwarf binaries.  They are interesting, both by
themselves, and as a way of explaining the fossil data found in white
dwarf -- neutron star binaries. The problem has been treated by
various authors, including Webbink, Rappaport, and Savonije
(\cite{WRS83}), who evolved such systems in the case of conservative
mass transfer from the red giant to the neutron star.

\subsection{Adiabatic Exponent: $\zs$} \label{seczs}

The adiabatic response of a star to mass loss has long been understood
(see, for example, Webbink (\cite{webbink85}) or Hjellming and
Webbink~(\cite{hw87}) for an overview), and on a simplistic level, is
as follows.  Stars with radiative envelopes (upper-main sequence
stars) contract in response to mass loss, and stars with convective
envelopes (lower-main sequence and Hayashi track stars) expand in
response to mass loss.  
The physics is as follows.

A star with a radiative envelope has a positive entropy gradient near
its surface. The density of the envelope material, if measured at a
constant pressure, decreases as one samples the envelope at
ever-increasing radii.  Thus, upon loss of the outer portion of the
envelope, the underlying material brought out of pressure equilibrium
expands, without quite filling the region from which material was
removed.  The star contracts on its dynamical timescale, in response
to mass loss.

A star with a convective envelope has a nearly constant entropy
profile, so the preceding analysis does not apply.  Instead, the
adiabatic response of a star with a convective envelope is determined
by the scalings among mass, radius, density, and pressure of the
isentropic material.  For most interesting cases, the star
is both energetically bound, and expands in response to mass loss.

Given the above physical arguements, the standard explanation of
mass-transfer stability is as follows.  A radiative star contracts
with mass loss and a convective star expands.  If a convective star
loses mass by Roche lobe overflow, it will expand with possible
instability if the Roche lobe does not expand fast enough. If a Roche
lobe-filling radiative star loses mass, it will shrink inside its
lobe (detach) and the mass transfer will be stable.

This analysis is of only a meagre and unsatisfactory kind
(Kelvin,\cite{kelvin}), as it treats stellar structure in only the
most simplistic way: convective vs. radiative envelope.  

One can quantify the response of a convective star by adopting some
analytic model for its structure, the simplest being an
isentropic polytrope.  This is a model in which the pressure $P$ and
density $\rho$ vary as

\be
P(r) = K\rho (r)^{1+1/n} \label{Prscaling}
\ee

\noindent and the constituent gas has an adiabatic exponent related to
the polytropic index through $\gamma = 1+1/n$.  Other slightly more
realistic cases include those with $\gamma \not = 1+1/n$,
applicable to radiative stars; composite polytropes, with different
polytropic indices for core and envelope; and centrally condensed
polytropes, which are polytropes with a point mass at the center.
These are all considered in a paper by Hjellming and Webbink
(\cite{hw87}).  We use the condensed polytropes, as they are simple,
fairly realistic models of red giant stars, which tend in the limit of
low envelope mass to (pointlike) proto-white dwarfs which are the
secondaries in the low mass-binary pulsar systems.

What follows is a brief treatmant of standard and condensed polytropes,
as applicable to the adiabatic response to mass loss.

Scaling arguements give $\zs$ for standard polytropes.  Pressure is an
energy density, and consequently scales as $GM^2R^{-4}$.  The density
scales as $MR^{-3}$.  The polytropic relation (Eq. (\ref{Prscaling})) 
immediately gives the scaling between $R$ and $M$.  Since the material
is isentropic, the variation of radius with mass loss is the same as
that given by the radius-mass relation of stars along this sequence.
Consequently, for polytropic stars of index $n$, 

\be 
\zs = \frac{n-1}{n-3}\,. \label{zspoly} 
\ee

\noindent In particular, $\zs=-1/3$ when $n=3/2$
($\gamma=5/3$).\footnote{This formula also reproduces the two
following results.  The gas giant planets ($n=1$) all have
approximately the same radius. Massive white dwarfs ($n=3$) are
dynamically unstable.}
%\footnote{Particularly interesting are the two cases
%$(a)$ $n=3$ ($\gamma =4/3$), with $M\sim R^0$, and marginal stability
%against gravitational collapse and $(b)$ $n=1$, where the length scale
%is independent of the mass, so that all pure $n=1$ polytropes have the
%same radius.}

The above approximation is fine towards the base of the red giant
branch, where the helium core is only a small fraction of the star's
mass.  It becomes increasingly poor as the core makes up increasingly
larger fractions of the star's mass, which happens when the star
climbs the red giant branch or loses its envelope to RLOF.
Mathematically speaking, the scaling law that lead to the formula for
$\zs$ is broken by the presence of another dimensionless variable, the
core mass fraction.

A far better approximation to red giant structure, and only slightly
more complex, is made by condensed polytropes, which model the helium core
as a central point mass (see, {\it e.g.:} Hjellming
and Webbink (\cite{hw87})).  Admittedly, this is a poor
approximation, as concerns the core.  However, the star's radius is
much greater than that of the core, so this is a good first-order
treatment.  Furthermore, differences between this approximation and
the actual structure occur primarily deep inside the star, while the star
responds to mass loss primarily near the surface, where the fractional
change in pressure is high.  Overlap between the two effects is negligible.

Analysis of the condensed polytropes requires integrating the
equation of stellar structure for isentropic matter (Lane-Emden
equation), to get a function $R(S, M_c, M)$, and differentiating $R$
at constant specific entropy $S$ (adiabatic requirement) and core mass
$M_c$ (no nuclear evolution over one sound crossing time).  In
general, the Lane-Emden equation is non-linear, and calculations must
be performed numerically.  The cases of $n=0$ and $n=1$ are linear and
quasi-analytic.  The case of $n=1$ is presented below, as a
nontrivial, analytic example, both for understanding, and because it
can be used as a check of other numeric calculations of $\zs$.

The $n=1$ Lane-Emden equation is (cf.: Clayton
(\cite{clayton68})): 

\be 
x^{-2}\frac{d}{dx} x^2 \frac{d}{dx} \phi(x) = -\phi(x)\,. 
\label{LE1} 
\ee 

Here, $x = r\sqrt{2\pi G/K} = r/\ell$ is the scaled radial coordinate
 and $\phi = \rho/\rho_c$ is the density, scaled to its central
 value\footnote{In general, $\rho = \rho_c 
\phi^n$.}.  The
 substitution $\phi(x) = x^{-1}u(x)$ allows solution by inspection:

\be
\phi(x) = \frac{\sin (x)}{x} \hspace{0.2in}x\in\left[0,\pi\right]. 
\label{LE1s}
\ee

The polytropic radius is set by the position of the first root of
$\phi$ and is therefore at $R = \pi \ell$.  Similarly, 
% $M(x) = -4\pi\ell^3 \rho_c x^2\partial_x\phi(x)$, so 
$M = 4 \pi^2 \rho_c \ell^3$.
Eq.~(\ref{LE1}) shows that the density ($\phi$) may be rescaled, without
affecting the length scale, so $R$ is independant of $M$, and $\zs=0$.

Alternately, the $n=1$ polytrope admits a length scale, $\ell$ which
depends only on specific entropy, so the polytrope's radius is
independant of the mass.

Generalising to condensed polytropes, Eq.~(\ref{LE1s}) suggests the
extension:
\be
\phi(x) = \frac{\sin (x+x_0)}{x}
\hspace{0.2in}x\in\left[0,\pi-x_0\right]. \label{LE1cs}
\ee
For this model, the stellar radius, stellar mass, and core mass are
\bea
R&=&(\pi-x_0)R_0 \,, \label{LE1cr} \\
M&=&(\pi-x_0)M_0 \,, \label{LE1cm} \\
M_c&=& M(x\rightarrow 0) \nonumber \\
   &=& M_0\sin(x_0) \,, \label{LE1cmc}
\eea
\noindent for some $M_0$ and $R_0$.  The core mass fraction 
\be
m = \frac{M_c}{M} = \frac{\sin (x_0)}{\pi -x_0} \label{LE1cmf}
\ee
\noindent is a monotonic function in $x_0$, and increases from $0$ to
$1$ as $x_0$ increases from $0$ to $\pi$.  As the core mass fraction
increases towards $1$, the polytrope's radius decreases from $\pi R_0$
to $0$, so the more condensed stars are also the smaller ones.

%The $R_0$ and $M_0$ of the previous example were not necessary for
%computing $\zs$.  There is a given amount of matter and it has a
%certain specific entropy, so $R_0$ and $M_0$ are both specified and
%independant.  In this example, $R_0$ is still irrelevant: $\zs$
%depends on $R$ only through its logarithmic derivative.  It is a
%different case with $M_0$, since two variables ($M_0$ and $x_0$) are
%needed to specify the two masses $M$ and $M_c$.  

The adiabatic $R-M$ exponent $\zs$ should be evaluated at constant
core mass, as opposed to mass fraction.  Thus, for condensed $n=1$
polytropes, 

\be 
\zs = \frac{\sin(x_0)}{\sin(x_0)+(\pi-x_0)\cos(x_0)}\,, \label{LE1czs} 
\ee

\noindent where $x_0$ is chosen to solve Eq.~(\ref{LE1cmf}).  This
solution matches that given by Eq.~(\ref{zspoly}) when there is no
core.  Furthermore, $\zs$ is an increasing function of the core mass
fraction, which diverges, as $m = M_c/M$ tends towards unity.  These are
general features of the condensed polytropes, and hold for polytropic
indices $n<3$.

%Calculation of $\zs$ for other values of the polytropic index must be
%carried out in some numeric analog of the above procedure.  Almost
%without exception, this will entail numerous integrations of the
%Lane-Emden equation for a variety of boundary conditions, and taking
%differences between close-by solutions.  Care must be taken in these
%calculations to make sure that derivatives are being taken at constant
%core mass and specific entropy.  

\begin{figure}
\centerline{\psfig{figure=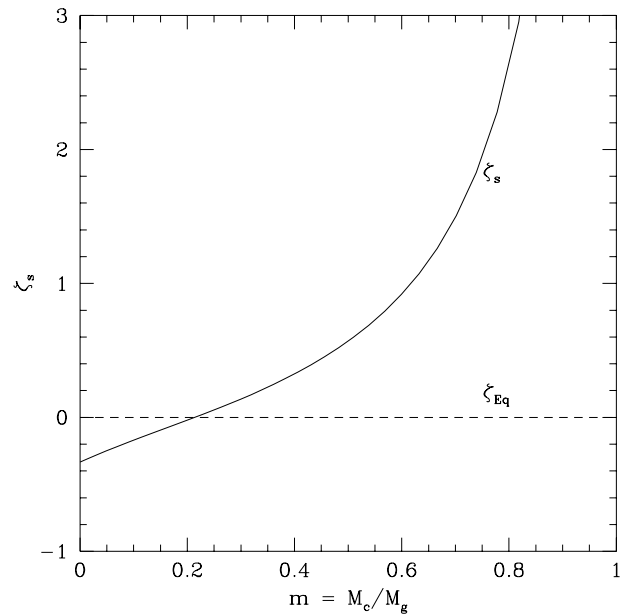,width=8.5cm,clip=t} {\hfil}}

\caption[] {Plot of $\zs$ versus core mass fraction, of an isentropic,
$n = 3/2$ red giant star.  As the fraction of mass in the core grows,
the star becomes less like a standard polytrope.  Important to note is
the crossing of the $\zs$ and $\zeq$ curves near $\mc/\mg = 0.2$.}
\label{zshwfig}
\end{figure}

\begin{figure}
\centerline{\psfig{figure=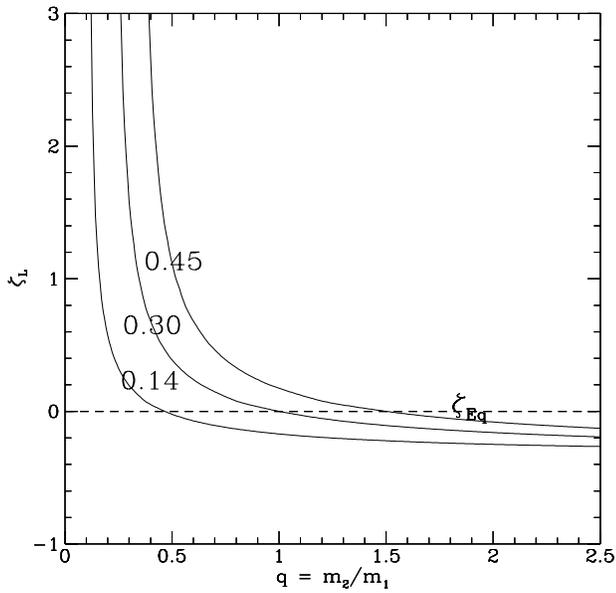,width=8.5cm,clip=t} {\hfil}}

\caption[] {Plots of $\zs$ versus q, assuming a fiducial $\mi = 1.4
\msun$ ($q = \mii/\mi = \mg/\mx = (\mc/\mi)(\mii/\mc) = $(const)/(core
mass fraction)).  The three solid-line curves correspond, in ascending
order, to core masses of $0.14$, $0.30$, and $0.45 \msun$.  One might
choose to photoenlarge this figure, as well as (\ref{zlextremes}),
et c., to use as overlays. }
\label{zshwvsq}
\end{figure}

The procedure for calculating $\zs$ is described in detail in
Hjellming and Webbink~(\cite{hw87}).  Results for a variety of core
mass fractions of $n=3/2$ polytropes are given, both in
Table~\ref{hwtable} and graphically, in Figs.~\ref{zshwfig}
and~\ref{zshwvsq}.  
 
\begin{figure}
\centerline{\psfig{figure=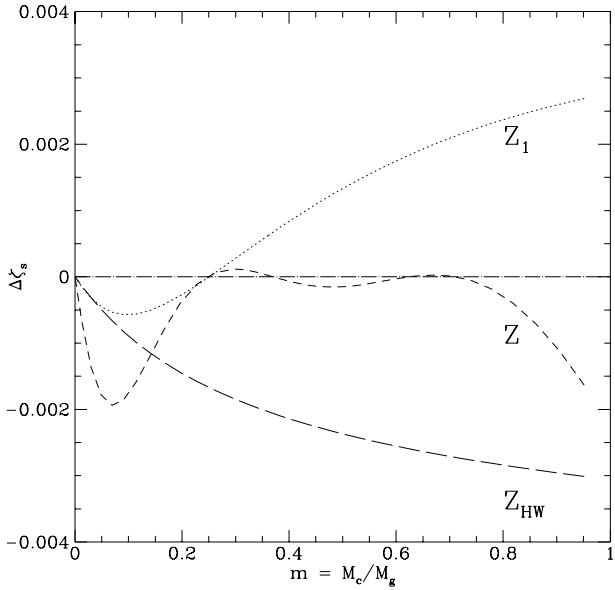,width=8.5cm,clip=t} {\hfil}}

\caption[] {Differences between various fit formulae and the function
$\zs(n=\frac{3}{2};m)$.  The labeled curves are as follows: $Z_{HW} =
0.01(\zshw-\zs)$, $Z_1=0.1(\zsspho-\zs)$, and
$Z=\zssph-\zs$.}
\label{residuals}
\end{figure}

The function $\zs(n=\frac{3}{2};m)$ can be reasonably well fit by the
function:

\be
\zshw = \frac{2}{3}\frac{m}{1-m} -\frac{1}{3} \label{zsfithw}
\ee
\noindent(Hjellming and Webbink,~\cite{hw87}), and to better than a
percent by either of the functions (in order of increasing accuracy):

\bea
\zsspho&=&\frac{2}{3}\left(\frac{m}{1-m}\right)
-\frac{1}{3}\left(\frac{1-m}{1+2m}\right)\,, \label{zsfitsph1} \\
\zssph & = & \zsspho -0.03 m +0.2\left[\frac{m}{1+\left(1-m\right)^{-6}}\right]\,,
\label{zsfitsph} 
\eea
\noindent as shown graphically in Fig. \ref{residuals}.

\begin{table}
\begin{tabular}{lrlr} \hline \hline 
\vspace{-2mm} \\ 
$E$	& $m_c$	 & $100\frac{dm_c}{dE}$  & $\zs$ \\ 
\vspace{-2mm} \\ \hline 
1.0	& 0.95144 & -4.7468   & 13.0293	\\
2.0 	& 0.90503 & -4.5372   & 6.31568	\\
3.0 	& 0.86066 & -4.3379   & 4.07567	\\
4.0 	& 0.81824 & -4.1484   & 2.95400	\\
5.0 	& 0.77766 & -3.9682   & 2.27964	\\
6.0 	& 0.73884 & -3.7967   & 1.82889	\\
7.0 	& 0.70170 & -3.6335   & 1.50588	\\
%8.0 		& 0.6661 & -3.478   & 1.2626	\\
%9.0 		& 0.6321 & -3.330   & 1.0726	\\
%10.0 		& 0.5995 & -3.189   & 0.9198	\\
%12.0		& 0.5383 & -2.927   & 0.6884	\\
%14.0   	& 0.4822 & -2.689   & 0.5206	\\
%16.0   	& 0.4306 & -2.472   & 0.3924	\\
%18.0   	& 0.3832 & -2.275   & 0.2903	\\
%20.0   	& 0.3395 & -2.096   & 0.2065	\\
%22.0   	& 0.2992 & -1.932   & 0.1358	\\
%24.0 		& 0.2621 & -1.783   & 0.0748	\\
%26.0   	& 0.2278 & -1.648   & 0.0211	\\
%28.0   	& 0.1961 & -1.524   & -0.0269	\\
%30.0   	& 0.1667 & -1.411   & -0.0706	\\
32.0   	& 0.13962 & -1.3079   & -0.11093	\\
34.0   	& 0.11442 & -1.2137   & -0.14849	\\
36.0    & 0.09101 & -1.1279   & -0.18390	\\
38.0    & 0.06925 & -1.0497   & -0.21759	\\
40.0    & 0.04898 & -0.9785   & -0.24991	\\
%42.0    	& 0.0300 & -0.9140  & -0.2811	\\
%44.0    	& 0.0123 & -0.8557  & -0.3114	\\
%45.4808 	& $3\cdot10^{-8}$  & -0.8166  & -0.3333 \\ 
\vspace{-2mm} \\ \hline 
\end{tabular}
\caption[] {Adiabatic $R-M$ relation {\it vs.} core mass fraction
$m_c$, as in Hjellming and Webbink~(\cite{hw87}), Table 3.  Columns
two and four are the core-mass fraction and mass-radius exponent,
respectively.  The parameter $E$ in columns one and three is an
alternate description of the degree of condensation of the polytrope,
used by Hjellming and Webbink.  The data presented here are in regions
where the residuals of the fit formula Eq. (\ref{zsfitsph}) are in
excess of $0.001$.}
\label{hwtable}
\end{table}

\subsection{Thermal Equilibrium Exponent: $\zeq$} \label{seczeq}

In the case of a red giants, fits to data consistently show that the
mass and luminosity depend almost entirely on the mass of the helium
core 
(Refsdal and Weigert, \cite{rw70}; cf.:
Verbunt~(\cite{verbunt93})).  Since the core mass changes on the
nuclear timescale, and we are interested in changes in the radius on
the (much shorter) thermal timescale, the radius may be taken as
fixed, giving $\zeq=0$.

\subsection{Roche Radius Exponent: $\zl$ } \label{seczl}

Since the exponent $\zl$ must be computed according to the evolution
of the binary with mass transfer, it is sensitive to mass transfer
mode, as are $\mt$, $a$, and $P$.  The results for the various modes
of nonconservative mass transfer are both interesting, and sometimes
counterintuitive.  It therefore makes sense to discuss them
systematically and at some length.

We rewrite $\zl$, in a form which depends explicitly on previously
calculated quantities:

\bea
\zl & = & \frac{\dln \rl}{\dln \mii} \nonumber \\
    & = & \frac{\dln a}{\dln \mii} + 
          \frac{\dln \rloa}{\dln q} 
	  \frac{\dln q}{\dln \mii}\; . \label{zl1}
\eea

\noindent The derivatives $\frac{\dln a}{\dln \mii}$ and $\frac{\dln
q}{\dln \mii}$ appear in Sect.~(\ref{secmx}), both for the unified
model, and for the ring; as well as in a tabulated form in Appendix
\ref{limits}.  All that remains is $\frac{\dln\rloa}{\dln q}$, which
will be calculated using Eggleton's (\cite{eggleton83}) formula for
the volume-equivalent Roche radius:

\bea
\rl/a & = & \frac{0.49 q^{2/3}}{0.6q^{2/3} + \ln(1+q^{1/3})}\; ,
\label{egg} \\
\frac{\dln \rloa}{\dln q} & = & \frac{q^{1/3}}{3} \times \nonumber \\
& & \left( \frac{2}{q^{1/3}}- 
\frac{1.2q^{1/3}+\frac{1}{1+q^{1/3}}}{0.6q^{2/3}+\ln(1+q^{1/3})}  
\right)
\; .\label{dregg}
\eea

\begin{figure}
\centerline{\psfig{figure=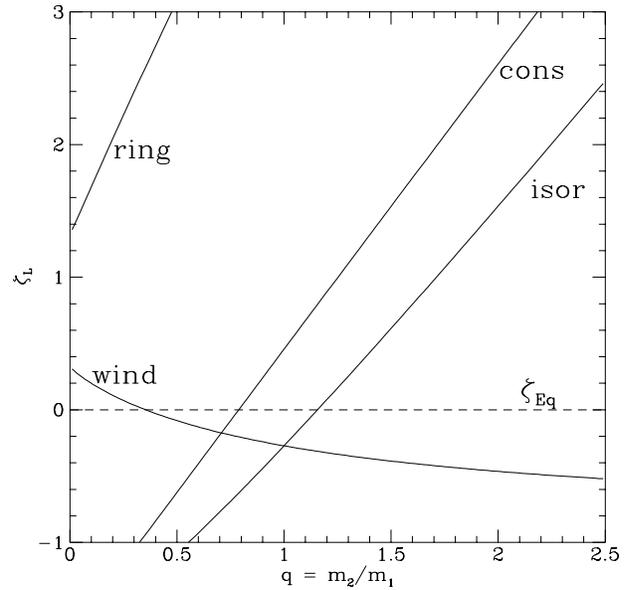,width=8.5cm,clip=t} {\hfil}}

\caption[] {$\zl$ with all mass transfer through a single channel.  For
each curve, all mass from the donor star is transferred according to
the indicated mode: conservative (cons); isotropic wind from donor
star (wind); isotropic re-emission of matter, from vicinity of
`accreting' star (iso-r); and ring formation, with $\gamma = 1.5$.
Since the unified model (winds+accretion) always has $0 \leq \alpha,
\beta, \alpha+\beta\leq1$, the $\zl(q)$ curves labeled cons, wind, and
iso-r also form an envelope around all curves in the unified model.}
\label{zlextremes}
\end{figure}

\begin{figure}
\centerline{\psfig{figure=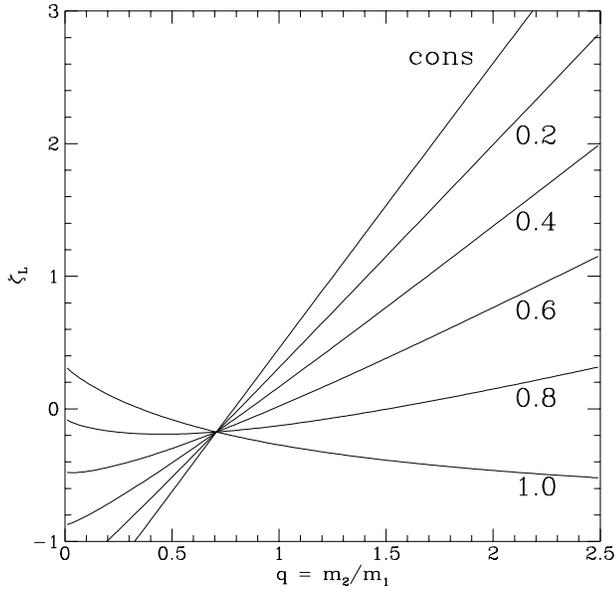,width=8.5cm,clip=t} {\hfil}}

\caption[] {This illustrates the effect of an increasing fraction of
the mass lost from the donor star ($\mii$) into a fast wind from that
star.  Solid lines correspond to $\alpha\in 0.0(0.2)1.0$ from top to
bottom on the graph's right side; $\alpha=0$ corresponds to
conservative mass transfer.  Note that at $q \sim 0.72$ ,
$\zl(\beta=0)$ is independent of $q$.}
\label{zlalpha}
\end{figure}

\begin{figure}
\centerline{\psfig{figure=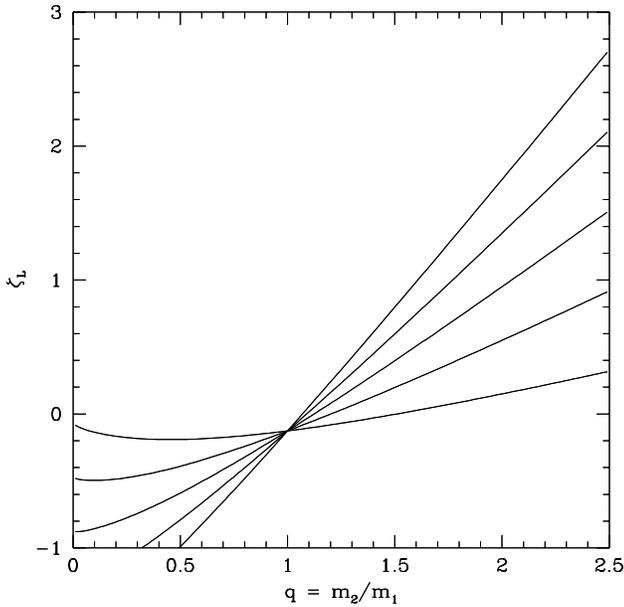,width=8.5cm,clip=t} {\hfil}}

\caption[] {The change in $\zl(q)$ for all mass lost to isotropic winds, with
varying fractions in isotropic wind and isotropically re-emiiited
wind.  From top to bottom on the right side, the solid curves are for
$(\alpha,\beta)$ of $(0.0,0.8)$, $(0.2,0.6)$, $(0.4,0.4)$,
$(0.6,0.2)$, and $(0.8,0.0)$.  Note that all curves cross at $q=1$.
Any set of curves $\alpha + \beta = const.$ will intersect at $q=1$,
as at this point, $\alpha$ and $\beta$ have the same coefficients in
$\zl$.}
\label{zladdsto08}
\end{figure}

\begin{figure}
\centerline{\psfig{figure=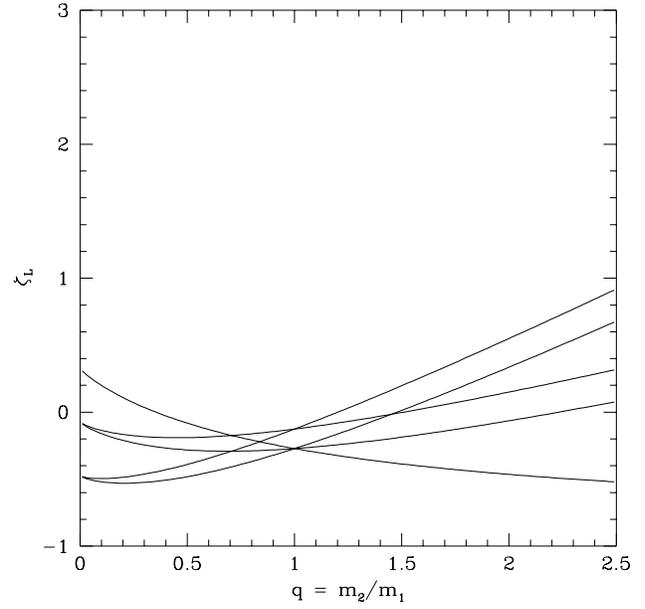,width=8.5cm,clip=t} {\hfil}}

\caption[] {$\zl$ for various $\alpha$ and $\beta$, so that most
transferred mass is ejected as winds.  From top to bottom on the right
side, the solid curves are for $(\alpha,\beta)$ of $(0.6,0.2)$,
$(0.6,0.4)$, $(0.8,0.0)$, $(0.8,0.2)$, and $(1.0,0.0)$. }
\label{zlvnoncon}
\end{figure}

To get an idea of how changing the mode of mass transfer effects
stability, $\zl$ has been plotted vs. $q$, for various models,
in Figs. ~\ref{zlextremes}, \ref{zlalpha}, \ref{zladdsto08}, and
\ref{zlvnoncon}.  

One notices several things in these graphs.  Figure ~\ref{zlextremes}
shows the extreme variation in $\zl(q)$ with changes in mode of mass
transfer.  Each of the three curves `ring', `wind', and `iso-r'
differs greatly from the conservative case. At least as important is
the extent to which they differ from one another, based only on the way in
which these three modes account for the variation of angular momentum
with mass loss.  When angular momentum is lost at an enhanced rate, as
when it is lost to a ring or a wind from the less massive star (direct wind,
at low-$q$; isotropic re-emission at high-$q$), the orbit quickly
shrinks in response to mass loss and $\zl$ is high.  By contrast, in the
case of the isotropic wind in the high-$q$ limit, angular momentum is
retained despite mass loss and the orbit stays wide, so $\zl$ is lower than
it is in the conservative case and the mass transfer is stabilised.

Second, Fig.~\ref{zlextremes} shows that ring formation leads to
rather high $\zl$, even for modest $\gamma^2=a_r/a=2.25$.  The formation of a
ring will usually lead to instability on the dynamical timescale ($\zs
< \zlr$).  The
slower thermal timescale instability will occur only if the giant is
rather evolved and has a high coremass fraction (with its high
$\zs$). 

Delivery of mass to a ring would imply either a substantially larger,
simultaneous flow of mass through the stabler of the two Jeans's
channels, or else orbital decay, leading to dynamical timescale mass
transfer.

Fig.~\ref{zladdsto08} shows the variation of $\zl$ due only to the
differences between the isotropic wind and isotropic re-emission, in
a family of curves with $\alpha + \beta = 0.8$.  At $q=1$, stability
is passed from the mostly isotropic re-emission modes at low $q$ to
mostly wind modes at high $q$.  The $q=1$ crossing of all these curves
is an artifact of the model.  Mass loss depends on the parameters in
the combination $\alpha + \beta$.  Angular momentum loss has a
dependence $A\alpha + \beta q^2$.  The two parameters $\alpha$ and
$\beta$ are then equivalent at $q = \sqrt{A} = 1$.

For $q<1$, mass transfer is stabilised by trading isotropic
re-emission for wind, so families of $\alpha+\beta = const$ curves
lie below their respective $(\beta=0, \alpha)$ curves in this region
of the $q$ -- $\zl$ graph.  This becomes interesting when one examines the
$\beta=0$ curves, at high $\alpha$.  For total wind
strengths of less than about 0.85 and $q<1$, $\zl < 0$.  Thus, for
modest levels of accretion (at least $\sim 15\%$), with the
remainder of mass transfer in winds, red giant - neutron star mass
transfer is stable on the thermal timescale, so long as $q < 1$.  If
the donor red giant has a modest mass core, so that $\zs > 0$, then
the process will be stable on he dynamical timescale, as well.

A family of curves with one wind of fixed strength and one wind of
varying strength (as shown in Fig. ~\ref{zlalpha}, with $\beta = 0$,
and variable $\alpha$), will also intersect at some value of $q$.  It
is easy to understand why this happens. For concreteness, take a
constant-$\beta$ family of curves.  Based on the linearity of the
equations in $\alpha$ and $\beta$, $\zl(q;\alpha,\beta) = f(q;\beta) +
\alpha g(q)$, with some $f$ and $g$.  If one evaluates $\zl(q)$ at a
root of $g(q)$, then the result is independent of $\alpha$.

Notice in each case, intersections of various unified wind model
$\zl(q)$ curves occur at $q = O(1)$.  One can understand this, in the
framework of the arguments of Sect.~(\ref{limits}).  Depending on
which side of $q=1$ one lies, either $\mi$ or $\mii$ has the majority
of the angular momentum.  When $q<1$, angular momentum will be lost
predominantly by the isotropic wind and $\zl$ will be high.  When
$q>1$, angular momentum losses will be modest and $\zl$ low. The
situation is reversed for isotropic re-emission.  The two curves
form a figure-X on the $q$ -- $\zl$ diagram and must cross when
niether $q$, nor $1/q$ is too great; that is, near $q=1$.

\section{An Example: Mass Transfer in Red Giant - Neutron Star Binaries}
\label{rgns}

Consider a binary system composed of a neutron star and a less evolved
star.  The less evolved star burns fuel, expanding and chemically
evolving.  If the binary orbital period is sufficiently short, the
evolving star will eventually fill its Roche lobe and transfer mass to
its companion.  We now consider this problem, in the case where mass
transfer starts while the donor is on the red giant branch (The
so-called case B.  See e.g.: Iben and Tutukov, ~\cite{IT85}.).

The global properties of an isolated star are functions of the stellar
mass and time from zero age main sequence.  Alternately, red giant
structure can be parameterised by total mass and core mass.  Detailed
models show that the dependence on total mass is weak,
so the radius and luminosity are nearly functions of the core mass,
only.  We use the fit formulae by Webbink (\cite{webbink75}), who
writes:

\bea
r & = & \rsun \exp( \sum c_{\imath} y^{\imath} ) \; , \label{rst} \\
L & = & \lsun \exp( \sum a_{\imath} y^{\imath} ) \label{lst} \\
  & = & X \cno \dot m_c \; . \label{dmc} 
\eea

\noindent Here $y = \ln(4 \mc / \msun)$, and the parameters
$a_{\imath}$, and $c_{\imath}$ are the result of fits to the red giant
models (See Table \ref{rgparams}.).  Equation (\ref{dmc})
assumes that the red giant's luminosity comes solely from shell
hydrogen burning by the {\sc CNO} cycle, which produces energy at a
rate $\cno = 5.987\cdot 10^{18} \mbox{erg g}^{-1}$ (Webbink,
et al., ~\cite{WRS83}).

\begin{table}
\begin{tabular}{r|cccc}
Z & $a_0$ & $a_1$ & $a_2$ & $a_3$ \\
\hline
0.02 & 3.50 & 8.11 & -0.61 & -2.13 \\
$10^{-4}$ & 3.27 & 5.15 & 4.03 & -7.06 
\end{tabular}
\vspace{3mm} \\
\begin{tabular}{r|cccc}
Z & $c_0$ & $c_1$ & $c_2$ & $c_3$ \\
\hline
0.02 & 2.53 & 5.10 & -0.05 & -1.71 \\
$10^{-4}$ & 2.02 & 2.94 & 2.39 & -3.89 
\end{tabular}
\caption[] { Parameters fitted to a series of red giant models, both for
Pop I ($Z = 0.02$, $X = 0.7$) and Pop II ($Z = 10^{-4}$, $X =
0.7$).  Data are transcribed from Webbink, {\it et. al.,}
(\cite{WRS83}), and are applicable over the ranges $y \in (-0.4, 0.6
)$ and $ y \in (-0.2, 0.4)$ for Pop I and Pop II, respectively. Pop I
figures due to Webbink (\cite{webbink75}); Pop II from Sweigart and
Gross (\cite{sg78}). }
\label{rgparams}
\end{table}

The red giant eventually fills its Roche lobe, transferring mass to
the other star.  If mass transfer is stable, according to the criteria
in Sect.~\ref{zetas}, then one can manipulate the time derivatives
of $r$ and $\rl$ to solve for the rate of mass loss from the
giant.  The stellar and Roche radii vary as:

\bea
\dot r & = & 
\left. \frac{\partial r}{\partial t}\right|_{\mg} 
+ 
r\; \zeq\frac{\dot \mg}{\mg} 
\; , \label{drst} \\
\dot \rl & = & 
\left. \frac{\partial\rl}{\partial t}\right|_{\mg} 
+ 
\rl \zl \frac{\dot \mg}{\mg}
\; . \label{drl2}
\eea

\noindent So long as the star remains in contact with its Roche lobe,

\bea
\frac{\dot \mg}{\mg} & = & \frac{1}{\zl - \zeq}
\left( 
\frac{\partial\ln r}{\partial t} - \frac{\partial\ln\rl}{\partial t} 
\right)_{\mg} \; , 
\label{dmg2} \\
\vspace{2mm} \nonumber \\
\frac{\dot r}{r} & = & \frac{\dot \rl}{\rl} \; , \nonumber \\
\frac{\dot r}{r} & = & \left. \frac{\partial r}{\partial t}\right|_{\mg} 
\left( \frac{\zl}{\zl-\zeq} \right) +
\left. \frac{\partial \rl}{\partial t}\right|_{\mg} 
\left( \frac{\zeq}{\zl-\zeq} \right) \; . \label{dlrdt2}
\eea

\noindent The first term in Eq. (\ref{drl2}) takes into account
changes in the Roche radius not due to mass transfer, such as tidal
locking of a diffuse star or orbital decay by gravitational wave
radiation.
For the models considered here, the Roche lobe evolves only due to
mass transfer. Equations~(\ref{drl2}), (\ref{dmg2}), and
(\ref{dlrdt2}) then reduce to the two equations:

\bea
\dot \rl & = & \rl \zl \frac{\dot \mg}{\mg} \; . \label{drl} \\
\frac{\dot \mg}{\mg} & = & \frac{1}{\zl - \zeq}\left( \frac{\partial\ln
r}{\partial t}\right)_{\mg} \; . \label{dmg} 
\eea

\noindent The additional relation:

\be
\frac{\dln r}{\partial t} = \frac{\dln r}{\dln\mc} 
\frac{\dln\mc}{\partial t} \;  \label{drdmc} 
\ee

\noindent is a consequence of the star being a red giant.  

The equations for the evolution of the core mass, red giant mass,
neutron star mass, orbital period, and semimajor axis (obtainable from
Eqs.~(\ref{dmc}), (\ref{dmg}), (\ref{drdmc}), (\ref{dmtdmw}),
(\ref{dPw}), and (\ref{daw})), form a complete system of first
order differential equations, governing the evolution of the red giant
and the binary.  The core mass grows, as burned hydrogen is added from
above.  This causes the star's radius to increase, which forces mass
transfer, increasing $a$ and $P$, at $\rl = r_g$.

\begin{figure}[h!]
\centerline{\psfig{figure=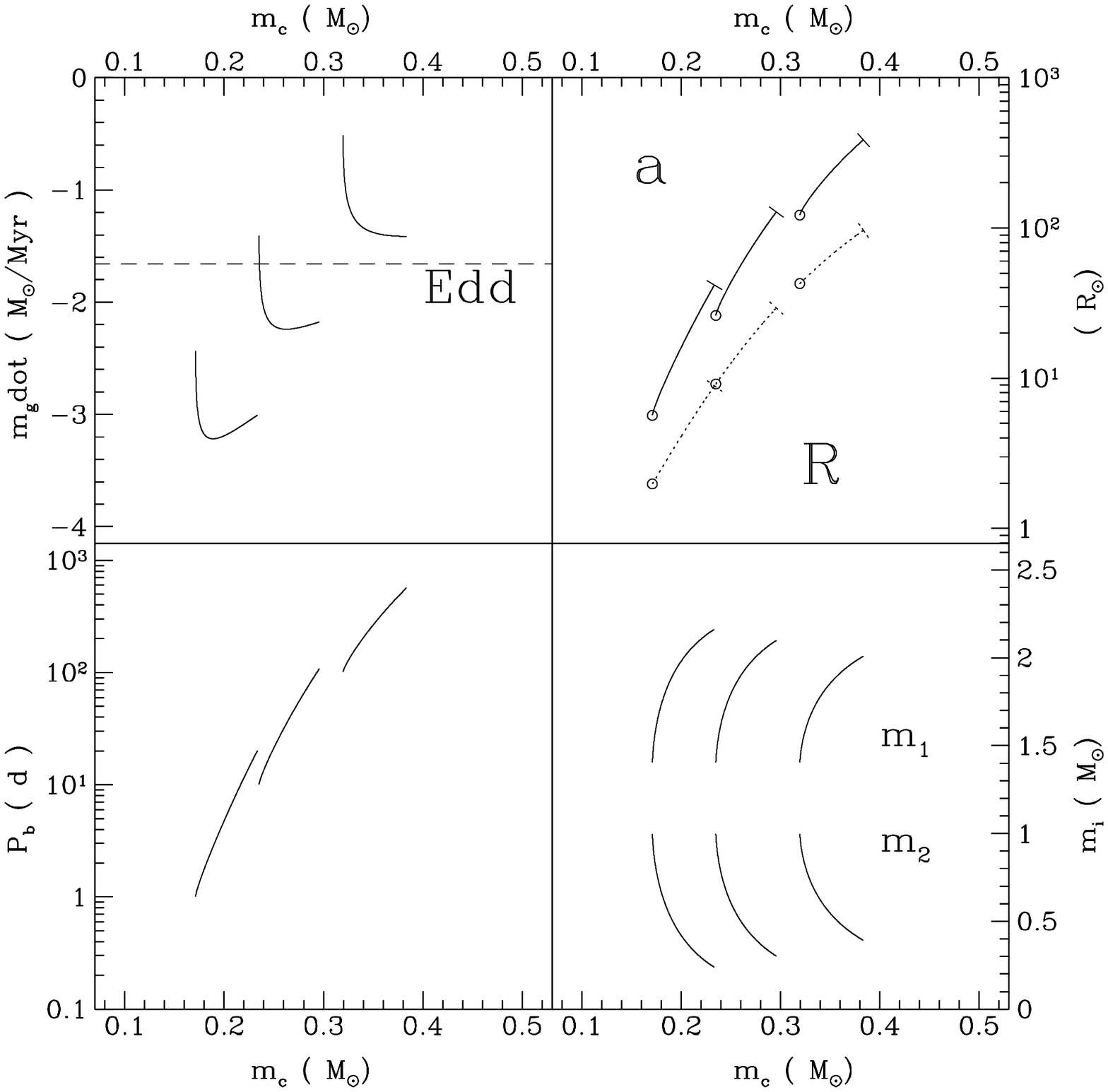,width=8.5cm,clip=t} {\hfil}}

\caption[] {This figure shows conservative evolution in three
different red giant neutron star binaries, initial with $\mx =
1.4\msun$, $\mg = 1.0 \msun$, and $P_i = 1, 10, 100$ days.  Red giant
coremass increases monotonically as the system evolves and has been
chosen as the independent variable.  The lower right hand panel shows
the evolution of the two stars' masses, with increasing $\mc(t)$;
the neutron star's mass ($\mi$) increases, with complementary a
decrease in the mass of the giant ($\mii$).

Orbital period evolution is shown in the lower left hand panel.  The
mass transfer decreases the mass of the lower mass star, and
consequently widens the orbit ($P$ increases, at constant $\mt$ and
$L$; Eq.~(\ref{Pw})).  The upper right hand panel shows the evolution
of semimajor axis (upper, solid curves) and red giant Roche radius
(lower, dashed curves).  The low and high coremass ends of each
segment in this panel are indicated by a circle and crossbar, for
clarity.  While transferring mass, the red giant fills its Roche lobe,
so the dashed segments shown here are consistent with the giant's
coremass radius relation (Eq.~(\ref{rst})).

Finally, the upper left hand panel shows the red giant mass loss
rate. The dotted line labeled `Edd' is the Eddington limit accretion
rate of the neutron star ($\dot\mx = 4\pi R_X c / \kappa_{Th}$, with
an assumed neutron star radius $R_X = 10\,km$ and $X_H =
0.70$. General relativistic effects and the variation of neutron star
mass and radius with accretion have been ignored.).  At high
coremasses, where the red giant's evolution is rapid, the mass loss
rate can be far in excess of the neutron star's Eddington rate,
implying that mass transfer is not always conservative.}
\label{xfercons}
\end{figure}

\begin{figure}[h!]
\centerline{\psfig{figure=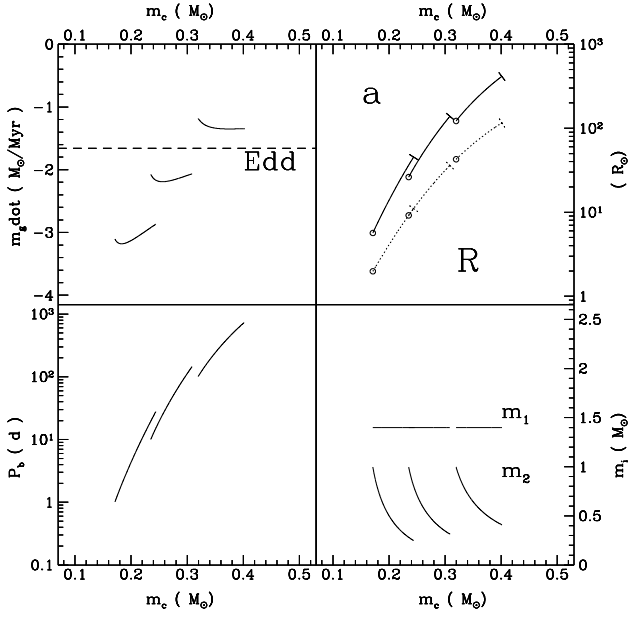,width=8.5cm,clip=t} {\hfil}}

\caption[] {Mass transfer by isotropic re-emission ($\alpha=0$,
$\beta=1$); details as in Fig.~\ref{xfercons}. }
\label{xferisorem}
\end{figure}

\begin{figure}[h!]
\centerline{\psfig{figure=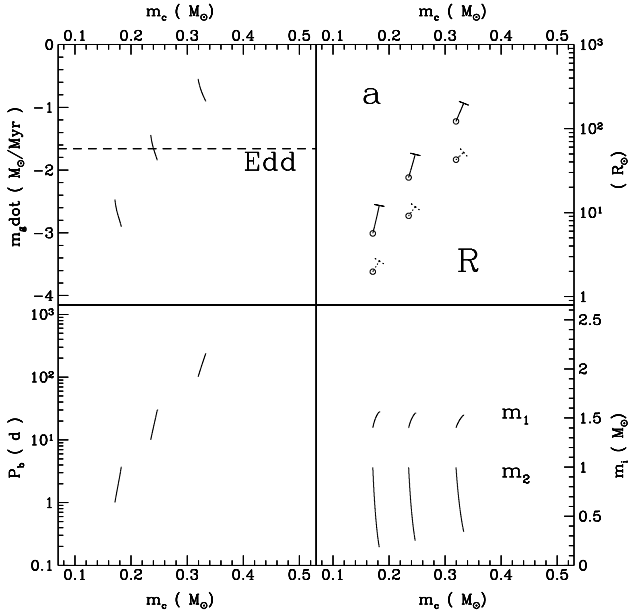,width=8.5cm,clip=t} {\hfil}}

\caption[]{ Mass transfer by a combination of accretion and wind
($\alpha=0.8$, $\beta=0$).  The pure wind gives similar results for
initial periods of 10 and 100 days, but leads to instability for short
initial periods.  This instability in low core mass systems is easily
understood, since shorter period systems can evolve towards lower mass
ratio systems, where $\zl(\alpha=1,\beta=0) > \zeq$.  See also
Fig. \ref{xfercons} for details. }
\label{xferwind08}
\end{figure}

\begin{figure}[h!]
\centerline{\psfig{figure=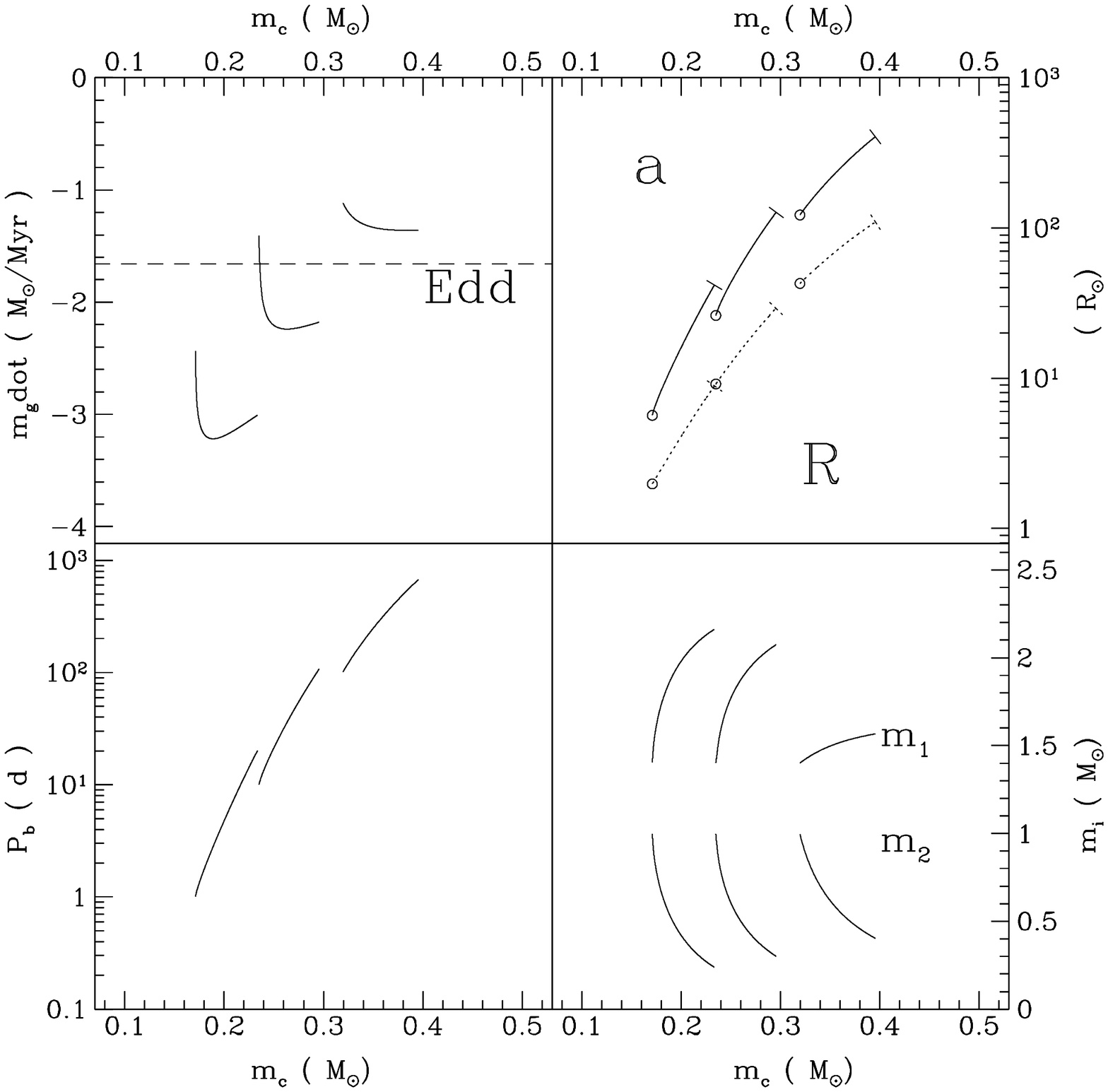,width=8.5cm,clip=t} {\hfil}}

\caption[] {Mass transfer with isotropic re-emission only so strong
as to insure $\dot\mx \leq \mxed$.  For details, see
Fig. ~\ref{xfercons}.}
\label{xfertoy}
\end{figure}

\subsection{The Code}\label{code}

The program used to generate the numbers presented here follows the
prescription outlined by Webbink, et al. ~\cite{WRS83}, in the
treatment of initial values, numeric integration of the evolution, and
prescription for termination of mass transfer.  The solar mass,
radius, and luminosity are taken from Stix (\cite{stix}).

Initial values were provided for the masses of the neutron star, the
red giant, and its core at the start of the contact phase (red giant
filling its Roche lobe): $\mx$, $\mg$, and $\mc$, respectively, as
well as for the parameters ($\alpha$, $\beta$, $\gamma$ ...) of the
mass transfer model.  A tidally-locked system was assumed and spin
angular momentum of the stars neglected.  The program then solved for
the orbital period $P$ and separation, $a$, using relations (\ref{kep}),
(\ref{egg}), and (\ref{rst}).

Integration of Eqs.~(\ref{dmidmw}), (\ref{dP}), (\ref{dmc}), and
(\ref{dmg}) was performed numerically by a fourth-order Runge-Kutta
scheme (cf.: Press, et al. \cite{nr}) with time steps
limited by $\Delta \mg \leq 0.003 \mg$, $\Delta \mc \leq 0.001 \mc$,
and $\Delta m_e \leq 0.003 m_e$.  The first two criteria are those
used by WRS; the last was added to this code, to insure that care is
taken when the envelope mass, $\me$, becomes small toward the end of
the integration.

Detailed numeric calculations (Taam, \cite{taam83}) show that a red
giant cannot support its envelope if $\me/\mg < 0.02$. The code
described here follows that of WRS and terminates mass transfer at
this point.

An overdetermined system of $\mc$, $\mg$, $\mx$, and $P$ was
integrated numerically by the code, allowing for consistancy checks of
the program.  Tests performed at the end of the evolution included
tests of Kepler's law (Eq.~(\ref{kep})), angular momentum
evolution (Eq.~(\ref{Pw}) or (\ref{Pr})), and a check of the
semi-detatched requirement $\rii = \rl = a \cdot \rloa$.
Plots of various quantities versus red giant core mass
show the evolutionary histories of binary systems, in Figs.
\ref{xfercons}, \ref{xferisorem}, and \ref{xferwind08}.  Population I
stars were used in all calculations (see Table ~\ref{rgparams}; $Z =
0.02$).

%asdf

Another interesting case of mass transfer is isotropic
re-emission at the minimum level necessary to ensure Eddington
limited accretion:

\be
\beta = \max(0, \dot\mg / \mxed -1) \, .\label{bedd}
\ee

Re-emission would presumably be in the form of propeller ejecta
(Ghosh and Lamb, (\cite{gl}) or a bipolar outflow, such as the jets
seen in the galactic superluminal sources. The evoution of a binary,
subject to this constraint, is shown in Fig.  \ref{xfertoy}.
Comparing Fig. \ref{xfertoy} with Figs. \ref{xfercons} and
\ref{xferisorem}, one sees that the low core mass systems, with their
corresponding low mass transfer rates, mimic conservative systems.
The faster evolving systems behave more like systems with pure
isotropic re-emission.

Almost all ring-forming systems are unstable to thermal and/or
dynamical timescale runaway of mass transfer, and are not displayed.

\section{Conclusion}

Previous works in the field of mass transfer in close binaries usually
centered on conservative mass transfer, even when observation directs
us to consider mass loss from the system, as is the case with white
dwarf neutron star binaries.  If these systems form from mass transfer
in red giant neutron star systems, then one must find a way to start
with a secondary star sufficiently massive to evolve off the main
sequence in a Hubble time, ($\mii > 1\msun$), in a binary with a $\mx >
1.3\msun$ neutron star and reduce the secondary's mass by $\approx 0.6
\msun$, while keeping the neutron star below $\sim 1.45 \msun$.  To do
this, some 3/4 of the mass from the secondary must be ejected from the
system.  Results here indicate that it is possible to remove this
much mass in winds, while maintaining stable mass transfer on the
nuclear timescale.

For the most part, nonconservative mass transfer, in which mass is
lost in fast winds, mimics the conservative case.  For the typical
initial mass ratios ($q \approx 0.7$), $\zl$ ranges from about -0.7
to -0.2.  The rate of mass transfer is given by equation (\ref{dmg}),
and is longest at the start of mass transfer, when $\mc$ is low. The
total time is therefore set almost entirely by $\mc(0)$ and $\zl(0)$,
and so differs by maybe a factor of 3, over all possible wind models.

It is worth repeating that changes in $\mt$ arising from
nonconservative evolution do not alter the relationship between white
dwarf mass and binary period by more than a couple of percent.  This
is true, by the following arguement.  The final red giant and core
masses differ only at the few percent level.  Approximately, then, the
red giant mass sets the red giant radius and therefore the Roche
radius.  In the (low $q$) approximation used by Paczy\'{n}ski
(\cite{paczynski71}), orbital period is a function of $\rl$ and
$\mii$, only.  $P$ vs.  $\mc$ is a function of the final state,
alone.  Using Eggleton's formula instead of Paczy\'{n}ski's introduces
only a very weak dependence on the mass of the other (neutron) star.
In the end, the theoretical motivation for the existance of a $P$ --
$\mc$ relation is significantly more solid than, say, our knowledge of
the red giant $R$ -- $\mc$ relation, on which the exact $P$ --
$\mc$ curve depends.

The exact mode of mass transfer will effect $P/P_0$, as is evident
from Eq.~(\ref{Pw}).  This could be important, in statistical studies
of white dwarf neutron star binaries, and trying to predict the
distribution of $P$ from the initial mass function, and distribution
of initial orbital periods.  This is dependant, of course, on the
development of a quantitative understanding of the common envelope
phase.

Finally, and probably the most useful thing, is that if one assumes
only accretion and wind-like mass transfer, then most every binary in
which mass is transferred from the less massive star is stable on both
dynamical and thermal timescales.  If the mass donor has a radiative
envelope (not treated here), it will shrink in response to mass loss,
and lose mass stably.  If the donor has a convective envelope, a
modestly sized core will stabilize it sufficiently to prevent mass
loss on the dynamical timescale.  Only if one has a very low mass (or
no) core, will the mass transfer be unstable on the dynamical
timescale, and then, only for $\alpha \rightarrow 1$.  High values of
$\alpha$, and low core mass in a convective star may lead to
instability on the thermal timescale, if the mass ratio, $q$ is
sufficiently low (see Fig. \ref{zlalpha}).

\begin{acknowledgements}
This work was supported by NSF grant AST93-15455 and NASA grant
NAG5-2756.
\end{acknowledgements}

\appendix

\section{Mass Transfer Models With Extreme Mass Ratios}\label{limits}

The results of Secs.~\ref{unified} and \ref{ring} are examined here in
the limiting cases of $q$ tending toward zero and infinity, to explore
the relations between the different modes, and the connections to
well-known toy models.  This is done to develop an intuition which may
be used in comparing the stability of mass transfer by the various
modes.

In considering extreme values of the mass ratio, one makes the reduced
mass approximation, regarding all the mass as residing in one star of
fixed position and all the angular momentum in the other, orbiting
star.  Errors are only of order $q$ (or $1/q$, if this is small).  One
might keep the Solar System in mind as a concrete example. The Sun has
all the mass, and the total angular momentum is well approximated by
Jupiter's orbital angular momentum.  The total mass $\mt\sim\msun$ and
the reduced mass $\mu \sim M_J$, with fractional errors $\sim
M_J/\msun \sim 10^{-3}$.

Retention of the factors $A$ (Eqs.~(\ref{dLdmw}), et seq. of
Sect.~\ref{unified}) and $\delta$ (Eq.  (\ref{dLdmr}), of
Sect.~\ref{ring}) allow for broader comparisons between the two
models. Formulae appropriate to the two extreme limits are in Table
\ref{limitstab}.  Columns 2 and 3 pertain to the wind models; 4 and 5
to the ring.

It is now straightforward to compare the two models of mass transfer
and loss in the extreme limits where they should be equivalent.
Comparisons will be made first in the test-mass is donor limit, where
the unified and ring models correspond exactly. Second is a treatment
of the opposite limit, where the two models give different but
reconcilable results.

First, the winds and ring models are equivalent in the $q=0$ limit,
upon identification of $\alpha$ with $\delta$ and $A$ with
$\gamma$\footnote{Actually, one must only identify the products of
mass fraction lost and relative specific angular momenta: $A\alpha$
and $\gamma\delta$.}.  This symmetry is not difficult to understand.
In the $q=0$ limit, only the direct isotropic wind and ring remove
angular momentum.  In this case, isotropic re-emission is an
isotropic wind from a stationary source, and accretion is always
conservative of both mass and angular momentum.  In each torquing case
(direct wind and ring), the ejected mass removes specific angular
momentum at an enhanced rate --- $A$ in the winds model, $\gamma$ in
the ring model.  Thus, it makes sense that wherever $A$ and $\gamma$
appear in the equations, they are in the products $A\alpha$ and
$\gamma\delta$, the rate of angular momentum loss per unit mass lost
from $\mii$.

More interesting is that, in the strict $q=0$ limit, the
parameters $\alpha$ and $\delta$ are found only in the combinations
$A\alpha$ and $\gamma\delta$.  The independance from other
combinations of parameters can be understood by examining the ratio

\be
\eta = \left. \frac{\dln\mt}{\dln (L/\mu)} \right|_{ev} \label{eta}\, , 
\ee

\noindent which tells the relative importance of angular momentum and
mass losses in the orbit's evolution.  For small donor mass, $\eta$ is
small and mass loss without angular momentum loss is unimportant.  The
strict mass loss term will be important only when the coefficient of
the $\dln(L/\mu)$ term, $A\alpha-1$ or $\gamma\delta-1$, is of order
$q$ or less.  This happens, for example, in the Jeans' mode of mass
loss ($q\ll 1$, $A=\alpha=1$), discussed below.

Even neglecting the questions of stability important to tidally
induced mass transfer, the situation is different when the
mass--losing star has all the mass and almost no angular momentum.
In this case, both mass and angular momentum loss are important. Since
the more massive body is the mass donor, a non-negligible fraction of
the total mass of the system may be ejected. Furthermore, the angular
momentum per reduced mass changes via isotropic re-emission
and ring formation.  Therefore, both mass and angular momentum loss
play a r\^{o}le in the dynamics. The test of relative importance is
$\eta$, which in the limit of large $q$, goes as $1/q$. Again, in the
case of extreme mass ratios, changes in $L/\mu$ dominate the course of
evolution.  As before, there are times when the coefficient of the
$\dln(L/\mu)$ term, $1-\alpha$ or $\gamma\delta+(1-\delta)$, is of
order $q$, or less.  In this case, above arguments fail and the strict
mass loss term $\dln\mt$ must be included.

The evolution of the angular momentum per reduced mass shows a
significant difference the winds and ring models in their $q \gg 1$
limit.  In the winds case, one may write $q(1-\alpha) =
q(1-\alpha-\beta) + q\beta$. The first part is the fractional rate of
increase of $\mu$, with loss of mass from the donor star.  The second
is the rate at which angular momentum is lost from the system, by
isotropic re-emission.  Since $q\gg 1$, the mass losing star is
nearly stationary and mass lost in a direct wind removes no angular
momentum.

The ring + accretion case is slightly different, but may be
interpreted similarly.  The term $q(1-\delta)$ replaces
$q(1-\alpha-\beta)$ as the fraction of mass accreted onto the first
star; i.e.: $\dln \mu/\dln\mii$.  The term $q\beta$ is replaced
by $q\gamma\delta$, as the rate of loss of angular momentum. Although
the donor star is stationary in this $q \gg 1$ ring case, angular
momentum is still lost.  This is due to the particular construction of
the ring model, in which the angular momentum removed is proportional
to the system's $L/\mu$, not to the specific angular momentum of
either body.  Thus, in the ring model, even as $h_2$ tends towards
zero with increasing $q$, ejected mass will carry away angular
momentum.  This is the final {\em caveat}: the ring model is just a
mechanism for the rapid removal of angular momentum from a binary
system, and one should keep this in mind, particularly when $q\gg 1$.

One might also notice for both models, that if angular momentum loss
is not overly efficient ($A$ and $\gamma$ not too much greater than
1), then mass loss from the test mass widens the orbit and mass loss
from the more massive star shrinks the orbit, just like in
conservative mass transfer.  Also, $\eta$ is not everywhere small.  In
particular, when $q$ is of order unity, $\eta$ is also and the
$\dln\mt$ term becomes as important as $\dln(L/\mu)$ in the equations
of motion.

\subsection{Jeans's Modes}\label{jm}

Often, one talks of the Jeans's mode of mass loss from a binary, in
which there is a fast, sperically symmetric loss of mass.  The Jeans'
mode has two limits.  One is a catastrophic and instantaneous loss of
mass, as in a supernova event (see van den Heuvel (\cite{vdh94}), for
a discussion).  
% flannery and van dn heuvel is prob. about the same/ better? A&A art.
In this case, if the orbit is initially circular and more than half
the total mass is lost in the explosion, the system unbinds.  This is
can be explained, based on energetics.  Initially, the system is
virialised with $<T> = -1/2<V>$.  The loss of mass does not change the
orbital velocities, so the kinetic energy per unit mass remains the
same.  The potential energy per unit mass is proportional to the total
mass, so if more than half the mass is lost, $E = T + V > 0$ and the
orbit is unbound.  A more detailed analysis may be done, and will give
ratios of initial to final orbital periods and semi-major axes, for
given initial to final mass ratios (Blaauw (\cite{blaauw}); Flannery
and van den Heuvel (\cite{flan})).

% The term ``Jeans mode'' is much confused in the literature ...

The equations in this paper will not give the `standard' Jeans
solution and unbound orbits.  Unbinding the orbit from elliptical to
hyperbolic requires $e > 1$, where $e=0$ has been
assumed from the outset.  The argument used in the preceding paragraph
seemingly necessitates the unbinding of the orbit with sufficient mass
loss, but it is not applicable here, as it also assumes a conservation
of mechanical energy.  Mechanical energy is not conserved in the above
calculations (section (\ref{unified}), for example), as the presence of
dissipative forces to damp $e$ to 0 have been assumed.

The other limit of Jeans's mode is mass loss by a fast wind, on a
timescale slow compared to the orbital period.  In this case, there is
no preferred orientation for the Runge-Lenz vector (direction of
semimajor axis in an eccentric orbit), and the orbit remains circular
throughout the mass loss, with $\mt a = const$.  This is the limit of
the Jeans mode which our calculations reproduce.  Jeans's mode is an
example of a degenerate case (mentioned above; here, $A\alpha =1$ ),
where $\dln(L/\mu) =0$.  In this case, one may simply apply equation
(\ref{a}), and see that $G\mt a = (L/\mu)^2 = const$.

\begin{table*}[bpt] %bottom of page, page of floats, top of page.
\begin{tabular}{ccccc}
& \multicolumn{2}{c}{Winds} & \multicolumn{2}{c}{Ring} \\
& $q\rightarrow 0$ & $q\rightarrow\infty$ & $q\rightarrow 0$ &
   $q\rightarrow\infty$ \\ 
\hline \\
\vspace{2mm} 
$\frac{\dln q}{\dln \mii}$ & $1$ &
   $q(1-\alpha-\beta)$ & $1$ & $q(1-\delta)$ \\ 
\vspace{2mm} 
$\frac{\dln \mt}{\dln\mii}$ & $0$ &
   $\alpha+\beta$ & $0$ & $\delta$ \\ 
\vspace{2mm} 
$\frac{\dln L/\mu}{\dln \mii}$ & $A\alpha-1$ & $q(1-\alpha)$ &
   $\gamma\delta -1$ & $q(\gamma\delta + (1-\delta))$ \\
\vspace{2mm} 
$\frac{\dln P}{\dln q}$ & $3(A\alpha-1)$ & $3\frac{1-\alpha}{1-\alpha-\beta}$ &
   $3(\gamma\delta -1)$ & $3\frac{\gamma\delta + (1-\delta)}{1-\delta}$ \\
\vspace{2mm} 
$\frac{\dln a}{\dln q}$ & $2(A\alpha-1)$ & $2\frac{1-\alpha}{1-\alpha-\beta}$ &
   $2(\gamma\delta -1)$ & $2\frac{\gamma\delta +
(1-\delta)}{1-\delta}$ \\
\vspace{2mm} \\
\hline \\
\vspace{2mm} 
$\frac{q}{q_0}$ & $\frac{\mii}{\miio}$ & $\frac{\mto}{\mt}$ &
$\frac{\mii}{\miio}$ & $\frac{\mto}{\mt}$ \\
\vspace{2mm}
$\frac{\mt}{\mto}$ & $1$ & $\frac{q_0}{q}$  & $1$ & $\frac{q_0}{q}$ \\
\vspace{2mm}
$\frac{(L/\mu)}{(L/\mu)_0}$ & $\left(\frac{q}{q_0}\right)^{A\alpha-1}$
& $\left( \frac{q}{q_0} \right)^{\frac{A\alpha-1}{1-\alpha-\beta}}$
& $\left( \frac{q}{q_0} \right)^{\gamma\delta-1}$
&
$\left(\frac{q}{q_0}\right)^{\frac{(\gamma\delta+(1-\delta)}{1-\delta}}$
\\
\vspace{2mm}
$\frac{P}{P_0}$ &
\multicolumn{4}{c}{$\left(\frac{(L/\mu)}{(L/\mu)_0}\right)^3$} \\
\vspace{2mm}
$\frac{a}{a_0}$ &
\multicolumn{4}{c}{$\left(\frac{(L/\mu)}{(L/\mu)_0}\right)^2$} \\
\end{tabular}
\caption[] { Formulae describing orbital evolution in the limits of
extreme mass ratios, $q = \mii/\mi \ll 1 \mbox{ and } q \gg 1$. }
\label{limitstab}
\end{table*}

\section{Extensions to models considered above} \label{hfpf}
% Appendix \ref{hfpf} is available via anonymous ftp from {\tt
% cdsarc.u-strasbg.fr} (130.79.128.5) or at {\tt
% http://cdsweb.u-strasbg.fr/Abstract.html}
In the interest of completeness, a five-parameter model of mass
transfer, combining winds from both stars, ring formation, and
accretion is presented.  The treatment in Sects.~\ref{secmx} and
\ref{seczl} is followed.  Modifications necessary for inclusion of other 
forms of angular momentum loss, such as $\dot L_{GR}$, due to
gravitational radiation reaction, are also discussed.

Construction of this model is straightforward, and results from the
inclusion of the various sinks of mass and angular momentum, due to
various processes. The nonconservative part of each model makes its
own contribution to the logarithmic derivatives of $\mt$ and $L$:

\bea
\epsilon & \equiv & 1 - \alpha -\beta- \delta \label{edef5} \;, \\
\dln \mt & = & (1-\epsilon)\frac{q}{1+q} \,
	\dln \mii \; , \label{dmt5} \\
\dln  L  & = & \left(\frac{A\alpha + \beta q^2}{1+q} + 
	\gamma\delta(1+q)\right)\,\dln\mii \; , \label{dL5}
\eea

\noindent where each of the variables retains its old meaning. 
Replacing the old definition of $\epsilon$ in Eq.~(\ref{edefw}) with
the new definition in Eq.~(\ref{edef5}) makes Eqs.~(\ref{dmidmw}) -
(\ref{dmudmw}) applicable to this model, as well.

\noindent Contributions from other evolutionary processes, such as
gravitational wave radiation reaction; realistically prescribed
stellar winds
% (which we might not be able to parameterise by this
simple prescription)
; tidal evolution, et c. can also be added.
Each will make its own contribution to the angular momentum and total
mass loss.  For example, orbital decay by gravitational radiation
reaction (Landau \& Lifshitz, \cite{landafshitz2}) can be included as
an other sink of angular momentum:

\be
- \left.\frac{\partial L}{\partial t}\right._{GR} 
	= \frac{32 G\mu}{5c^5}(G\mt)^6(L/\mu)^{-8} \;. \label{GRtorque}
\ee

In most cases, this type of physics can be modelled as an intrinsic
$\dot \rl$ (the second term in Eq.~(\ref{drl2})).  Like the intrinsic
stellar expansion term of Eq.~(\ref{drst}), this kind of evolution
occurs even in the absence of mass transfer.  Therefore, the
convenient change of variables from $t$ to $q$ used in
Sect.~(\ref{secmx}) introduces singularities when evolution takes
place in the absence of mass transfer.  The equations of
Sect.~(\ref{secmx}) still hold, but only for those phases of the
binary's evolution durring which tidally-driven mass transfer takes
place.

We temporarily neglect these complications and consider mass transfer
via isotropic wind, isotropic re-emission, and formation of a ring,
with mass fractions $\alpha$, $\beta$, and $\delta$ respectively.  The
remainder of the mass transfer (the fraction $\epsilon =1-\alpha -
\beta - \delta$) goes into accretion.  The ratio $h_r/h_{bin}$ is
$\gamma$, where $\gamma^2 = a_{r}/a$.  Note that $\delta$, $\gamma$,
$\beta$, and $\alpha$ are all used as before; $\epsilon$ should still
be regarded as the accreted fraction.

\bea
\frac{\mt}{\mto} & = & \left( \frac{1+q}{1+q_0} \right)
	\left( \frac{1+\epsilon q_0}{1+\epsilon q} \right)
	\; , \label{m5} \\
\frac{a}{a_0} & = & 
\left(\frac{q}{q_0}\right)^{2{\cal A}-2}
\left(\frac{1+q}{1+q_0}\right)^{1-2{\cal B} }\nonumber
\\
& & \left(\frac{1+\epsilon q}{1+\epsilon q_0}\right)^
{3+2{\cal C}}
\label{a5} \\
\frac{P}{P_0} & = & 
\left(\frac{q}{q_0}\right)^{3{\cal A}-3}
\left(\frac{1+q}{1+q_0}\right)^{1-3 {\cal B}}\nonumber
\\
& & \left(\frac{1+\epsilon q}{1+\epsilon q_0}\right)^
{5+3{\cal C}}  
\label{P5} \\
\ca{A}{5} & = & A\alpha+\gamma\delta \label{qexp5} \\
\ca{B}{5} & = & \frac{A\alpha+\beta}{1-\epsilon} \label{qpoexp5} \\
\ca{C}{5} & = & \frac{\gamma\delta(1-\epsilon)}{\epsilon}+
\frac{A\alpha\epsilon}{1-\epsilon}+\frac{\beta}{\epsilon(1-\epsilon)} 
\label{eqpoexp5}
\eea

\noindent Taking $\alpha$, $\beta$ = 0, gives formulae for a ring of
strength $\delta$:

\bea
\frac{a}{a_0} & = & \left(\frac{q_0}{q}\right)^{2(1-\ca{A}{r})} 
\left(\frac{1+q}{1+q_0}\right) \nonumber \\ 
& & \times \left(\frac{1+(1-\delta)q}{1+(1-\delta)q_0}\right)^{3 +
2\ca{C}{r}} \label{ard} \\
\frac{P}{P_0} & = & \left(\frac{q_0}{q}\right)^{3(1-\ca{A}{r})}
	     \left(\frac{1+q}{1+q_0}\right) \nonumber \\
& & \times \left(\frac{1+(1-\delta)q}{1+(1-\delta)q_0}\right)^{5 +
3\ca{C}{r}} \label{Prd} \\
\frac{\dln a}{\dln q} & = & 2(\ca{A}{r}-1) + 
(1-2\ca{B}{r})\frac{q}{1+q} \nonumber \\ 
& & + (3+2\ca{C}{r}) \frac{q}{1+\epsilon q}
\, , \label{dard} \\
\frac{\dln P}{\dln q} & = & 3(\ca{A}{r}-1) + 
(1-3\ca{B}{r})\frac{q}{1+q} \nonumber \\ 
& & + (5+3\ca{C}{r})\frac{q}{1+\epsilon q} 
\, .\label{dPrd}
\eea
\noindent Where the relevant exponents are functions of the parameters
$\gamma$ and $\delta$:

\bea 
\ca{A}{r} & = & \gamma\delta \label{qexprd} \\ 
\ca{B}{r} & = & 0 \label{qpoexprd} \\ 
\ca{C}{r} & = & \frac{\gamma\delta^2}{1-\delta} \label{eqpoexprd} \\
\epsilon_r & = & 1 - \delta  \;.
\eea

It is also instructive to examine the model in the degenerate cases of
$\epsilon=0$ and $\epsilon=1$, where the functional forms change.
When there is no accretion ($\epsilon=0$), the standard $\ca{C}{5}$
becomes singular, while the term $1+\epsilon q$ approaches $1$.
Defining the singular part of $\ca{C}{5}$:

\bea
\ca{C}{sing} & = & \lim_{\epsilon\rightarrow 0} 
\epsilon\ca{C}{5} \nonumber \\
& = & \beta + \gamma\delta \; ,\label{cs}
\eea

\noindent the equations governing binary evolution can be rewritten, for the
case when no material is accreted:

% e = 0 (.eo) formulae:
\bea
\frac{L}{L_0} & = & \left(\frac{q}{q_0}\right)^{\ca{A}{5}} 
\left(\frac{1+q_0}{1+q}\right)^{\ca{B}{5}} \nonumber \\
& & \exp{[(q-q_0)\ca{C}{sing}]} 
\; , \label{Leo} \\
\vspace{2mm}
\frac{a}{a_0} & = & \left(\frac{q}{q_0}\right)^{2(\ca{A}{5}-1)} 
\left(\frac{1+q_0}{1+q}\right)^{1-2\ca{B}{5}} \nonumber \\
& & \exp{[2\ca{C}{sing}(q-q_0)]} 
\; , \label{aeo} \\
\vspace{2mm}
\frac{P}{P_0} & = & \left(\frac{q}{q_0}\right)^{3(\ca{A}{5}-1)} 
\left(\frac{1+q_0}{1+q}\right)^{1-3\ca{B}{5}} \nonumber \\
& & \exp{[3\ca{C}{sing}(q-q_0)]} 
\; , \label{Peo} \\
\vspace{2mm}
\frac{\dln a}{\dln q} & = & 2(\ca{A}{5}-1) + 
(1-2\ca{B}{5})\frac{q}{1+q} \nonumber \\ 
& & + (3+2\ca{C}{sing})q
\; , \label{daeo} \\
\vspace{2mm}
\frac{\dln P}{\dln q} & = & 3(\ca{A}{5}-1) + 
(1-3\ca{B}{5})\frac{q}{1+q} \nonumber \\ 
& & + (5+3\ca{C}{sing})q 
\; .\label{dPeo} 
\eea

It might be worth noting that the above considerations are irrelevant
for the pure $\alpha=1$ models, where $\beta$, $\delta$, and
$\epsilon$ vanish.  There is no profound reason for this.

In the case where all material is accreted ($\epsilon=1$), there are
seeming singularities in the coeficients $\ca{B}{5}$ and $\ca{C}{5}$.
Proper solution of the equations of evolution in this case, or setting
$\epsilon q \rightarrow q$ before taking limiting values of the
coeficients $\ca{B}{5}$ and $\ca{C}{5}$, shows that there is no
problem at all.  In this case, the equations reduce to Eqs.~(\ref{ac}),
(\ref{Pc}), and:

% e = 1(unity) (.eu) formulae:
\bea
\mt & = & \mto \; , \label{mteu} \\
\vspace{2mm}
L & = & L_0 \; , \label{Leu} \\
\vspace{2mm}
\frac{\dln a}{\dln q} & = & 4 \frac{q}{1+q} - 2 \label{daeu} \; ,\\
\vspace{2mm}
\frac{\dln P}{\dln q} & = & 6 \frac{q}{1+q} - 3 \label{dPeu} \; .
\eea

\begin{table}
\begin{tabular}{rccc}
& winds & ring & combined \\
\vspace{2mm} \\
$-\frac{\partial m_{mode}}{\partial \mii}$ & $\alpha ,\beta$ &
$\delta$ & $\alpha , \beta , \delta$ \\
$\epsilon = -\frac{\partial \mi}{\partial \mii}$ & $1-\alpha-\beta$ &
$1-\delta$ & $1-\alpha-\beta -\delta$ \\
\vspace{0mm} \\ \hline
\vspace{-2mm} \\
${\cal A}$&Eq.~(\ref{qexpw}) &Eq.~(\ref{qexprd}) &Eq.~(\ref{qexp5})  \\
${\cal B}$&Eq.~(\ref{qpoexpw}) &Eq.~(\ref{qpoexprd}) &Eq.~(\ref{qpoexp5})  \\
${\cal C}$&Eq.~(\ref{eqpoexpw}) &Eq.~(\ref{eqpoexprd}) &Eq.~(\ref{eqpoexp5}) 
\vspace{2mm} \\ \hline
\vspace{1mm} \\
\multicolumn{4}{c}{$\frac{\mt}{\mto} =
\left(\frac{1+q}{1+q_0}\right)\left(\frac{1+\epsilon q}{1+\epsilon
q_0}\right)^{-1}$\hspace{0.5cm}(Eq.~(\ref{mtw}))} \\
\vspace{2mm} \\
\multicolumn{4}{c}{$\frac{L/\mu}{L_0/\mu_0} =
\left(\frac{q}{q_0}\right)^{{\cal A}-1}
\left(\frac{1+q}{1+q_0}\right)^{1-{\cal B}}
\left(\frac{1+\epsilon q}{1+\epsilon q_0}\right)^{{\cal C} +1}$} \\
\vspace{2mm} \\
\multicolumn{4}{c}{$\frac{P}{P_0} =
\left(\frac{q}{q_0}\right)^{3{\cal A} -3}
\left(\frac{1+q}{1+q_0}\right)^{1-3{\cal B}}
\left(\frac{1+ \epsilon q}{1+\epsilon q_0}\right)^{3{\cal C}+5}$} \\
\vspace{2mm} \\
\multicolumn{4}{c}{$\frac{a}{a_0} =
\left(\frac{q}{q_0}\right)^{2{\cal A}-2}
\left(\frac{1+q}{1+q_0}\right)^{1-2{\cal B}}
\left(\frac{1+\epsilon q}{1+\epsilon q_0}\right)^{2{\cal C} +3}$} \\
\vspace{2mm} \\
\multicolumn{4}{c}{$\frac{\dln{a}}{\dln{q}} = 2({\cal A}-1) +
(1-2{\cal B})\frac{q}{1+q} + (2{\cal C} +3)\frac{\epsilon q}
{1+\epsilon q} $} \\
\vspace{2mm} \\
\multicolumn{4}{c}{$\frac{\dln{P}}{\dln{q}} = 3({\cal A}-1) +
(1-3{\cal B})\frac{q}{1+q} + (3{\cal C} +5)\frac{\epsilon q}
{1+\epsilon q} $} \\
\end{tabular}
\label{cheatsheet}
\caption []{This reference table is divided into three parts.  First
are the model parameter definitions.  An index of equations for the
coeficients ${\cal A}$, ${\cal B}$, and ${\cal C}$, relevant to each
particular model, follows.  Last are the various formulae of orbital
evolution derived in this paper.}
\end{table}
\end{document}